\pdfoutput=1
\documentclass{aa} 

\usepackage{graphicx}
\usepackage{amsmath,amssymb}
\usepackage[varg]{txfonts}
\usepackage{hyperref}
\hypersetup{
  pdftitle={Forward modeling of galaxy kinematics in slitless spectroscopy},
  pdfauthor={Mehdi Outini, Yannick Copin},
}
\usepackage{siunitx}            
\renewcommand{\d}{\ensuremath{\mathrm{d}}}

\newcommand{\Ha}{\ensuremath{\element{H}\alpha}}

\usepackage[normalem]{ulem}

\begin{document}
\title{Forward modeling of galaxy kinematics in slitless spectroscopy}

\author{M. Outini\inst{1}\and Y. Copin\inst{1}}

\institute{%
  Université de Lyon, F-69622, Lyon, France; Université de Lyon 1,
  Villeurbanne; CNRS/IN2P3, Institut de Physique Nucléaire de Lyon.
  \email{y.copin@ipnl.in2p3.fr} 
}

\date{Received <date> / Accepted <date>}


\abstract{%
  Slitless spectroscopy has long been considered as a complicated and confused
  technique. Nonetheless, with the advent of Hubble Space Telescope
    (HST) instruments characterized by a low sky background level and a high
    spatial resolution (most notably WFC3), slitless spectroscopy has become
  an adopted survey tool to study galaxy evolution from space. }{%
  We investigate its application to single object studies to measure not only
  redshift and integrated spectral features, but also spatially resolved
  quantities such as galaxy kinematics. }{%
  We build a complete forward model to be quantitatively compared to actual
  slitless observations.  This model depends on a simplified thin cold disk
  galaxy description including flux distribution, intrinsic spectrum and
  kinematic parameters -- and on the instrumental signature.  It is used to
  improve redshifts and constrain basic rotation curve parameters, i.e.\
  plateau velocity $v_{0}$ (in \si{km.s^{-1}}) and central velocity gradient
  $w_{0}$ (in \si{km.s^{-1}.arcsec^{-1}}).}{%
  The model is tested on selected observations from 3D-HST and GLASS surveys,
  to estimate redshift and kinematic parameters on several galaxies measured
  with one or more roll angles. }{%
  Our forward approach allows to mitigate the self-contamination effect, a
  primary drawback of slitless spectroscopy, and therefore has the
    potential to increase precision on redshifts.  In a limited
  sample of well-resolved spiral galaxies from HST surveys, it is possible to
  significantly constrain galaxy rotation curve parameters. This
  proof-of-concept work is promising for future large slitless spectroscopic
  surveys such as EUCLID and WFIRST.}

\keywords{%
    Galaxies: high-redshift -
    Galaxies: kinematics and dynamics -
    Galaxies: evolution -
    Instrumentation: spectrographs
}

\maketitle
%

\section{Introduction}
\label{sec:introduction}

Spectroscopy surveys play a fundamental role in the understanding of galaxy
formation and evolution with cosmic time and in cosmology. These surveys have
been achieved using different techniques.  Fiber-fed multi-object spectrographs
are commonly used to measure redshift and integrated properties on pre-selected
targets.  For instance, the Sloan Digital Sky Survey (SDSS) has measured more
than a million of local galaxy spectra \citep{york_sloan_2000,
  strauss_spectroscopic_2002}, and has revealed how the star formation rates,
metallicities, stellar populations vary with environment, mass and redshift
\citep{gomez_galaxy_2003, brinchmann_physical_2004,
  kauffmann_environmental_2004}. Similar surveys like the 2dF and 6dF Galaxy
Redshift Surveys \citep{folkes_2df_1999, jones_6df_2004} have also allowed to
constrain cosmological parameters by mapping the distribution of galaxies along
cosmic time \citep{cole_2df_2005}. More recently, the VIPERS survey
\citep{de_la_torre_vimos_2013} allowed to map with an unprecedented precision
the large-scale distribution of galaxies by measuring more than \num{100 000}
redshifts at $0.5<z<1.2$.  Nevertheless, these fiber-based surveys suffer from
drawbacks: galaxy central regions are integrated so two-dimensional internal
structure cannot be properly recovered, and objects need to be selected and
targeted \emph{a priori}.

Alternatively, Integral-Field Spectrographs (IFS) are ideal to study resolved
objects over their spatial extent, but still require explicit pointing of
individual galaxies.  In the last years, the development of highly successful
IFS surveys ($\mathcal{R} \gtrsim 1500$) -- such as SAURON
\citep{de_zeeuw_sauron_2002}, ATLAS$^{\text{3D}}$
\citep{cappellari_atlas3d_2011}, SAMI \citep{croom_sydney-aao_2012}, CALIFA
\citep{sanchez_califa_2012}, and MaNGA \citep{bundy_overview_2015} -- has
pushed our understanding of galaxy properties further.  Not limited to
integrated measurements, these surveys in the nearby Universe
($z \lesssim 0.1$) allowed to accurately map the gas, stellar populations and
kinematics, and led to a new kinematical classification scheme of early-type
galaxies \citep{emsellem_sauron_2007, emsellem_atlas3d_2011}.

Contrary to inherently targeted multi-object spectrographs, ``panoramic'' IFS
can be used to carry on \emph{un}targeted surveys, but only on very limited sky
areas because of observation time/cost constraints.  E.g., MUSE
\citep{bacon_muse_2010} has probed the evolution of gas kinematics of low-mass
galaxies ($M_{\star} \leq \SI{e10}{M_{\odot}}$) up to $z=1.4$ on the Hubble
Deep Field South \citep{contini_deep_2016}.

In contrast to graceful IFS, slitless spectroscopy has generally been seen as a
clumsy technique with well recognized drawbacks.  As a matter of fact, the
absence of an independent spatial sampling before spectral dispersion induces
two major contamination issues: self-contamination -- the effective spectral
resolution is directly related to the size and shape of the spatially resolved
object in the dispersion direction -- and cross-contamination -- signal
pollution from nearby objects; both effects make the data reduction difficult
and the redshift measurement less accurate.  In addition, slitless spectroscopy
is affected by a comparatively high integrated background level
(particularly from the ground), limiting the depth or SNR of
observations. However, this technique has some pros of its own: the ease of instrumental
design and observational use, a large field of view and a very high
multiplexing capability, all leading to very large object catalogs
($\sim \num{100 000}$ galaxies in recent surveys and $\gtrsim \si{15}{\mega}$
in future surveys). Furthermore, it has the potential to provide flux-limited
surveys with high spectro-photometric accuracy, insensitive to fiber or slit
losses.

Thereby, since the advent of the Hubble Space Telescope (HST) grism
instruments, slitless spectroscopy has \emph{de facto} become a tool of choice
to study galaxy evolution from space: a low background level and a fine spatial
resolution both mitigate the aforementioned shortcomings
\citep{freudling_hubble_2008, kummel_hubble_2011, dressel_wide_2012}.
Dedicated HST surveys such as WISP \citep{atek_wfc3_2010}, 3D-HST
\citep{brammer_3d-hst:_2012, momcheva_3d-hst_2016}, GLASS
\citep{schmidt_through_2014, treu_grism_2015}, FIGS
\citep{pirzkal_figsfaint_2017} have led to consider this technique as
appropriate to derive redshift and integrated galaxy properties over large
samples, and ready to be used in future missions as EUCLID
\citep{grupp_optical_2012} and WFIRST \citep{spergel_wide-field_2015}.

Traditional approaches in slitless spectroscopy
\citep[e.g.][]{kummel_slitless_2009} use standard ``inverse'' data reduction
and analysis methods, extracting parameters from observations using successive
and dedicated data manipulation steps.  Typically, it involves an empirical
modeling of the spectral trace, a cross-dispersion summation to estimate the 1D
galaxy spectrum, the \emph{ad-hoc} combination of spectra obtained at different
position angles, and any subsequent spectral analyses performed on the averaged
spectrum.  Not only the proper error propagation is difficult between the
different data-reduction steps, but such reverse approach can hardly correct
for or quantify the impact of spatially resolved galactic properties, such as
internal dynamics or metallicity gradients.

Alternatively, a ``forward'' approach allows to constrain physical or
instrumental parameters directly in the observation space, properly accounting
for degeneracy and covariances, and allowing for the inclusion of bayesian-like
priors.  By constructing a predictive model of the galaxy 2D dispersed image
(hereafter coined \emph{spectrogram}) depending on a set of observationally or
physically motivated parameters, we investigate the possibility to measure not
only intrinsic mean spectral quantities -- e.g., redshift, emission line
intensities and widths -- independently of self-confusion, but also spatially
resolved quantities such as internal kinematics.  Thus, by combining forward
methods to derive resolved quantities on single object and the large
multiplexing power of multi-object spectrographs, slitless spectroscopy surveys
offer a unique opportunity to study galaxy properties at an unprecedented
scale.

In this paper, we will detail how to forward-model slitless spectrograms from a
galaxy model -- including flux distribution, intrinsic spectrum and kinematic
parameters -- and an instrumental signature.  Considering our targets are
mainly line-emitting disk galaxies, we will use two majors assumptions: an
axi-symmetric thin cold disk geometry for the galaxy, and a separability
hypothesis under which the intrinsic galaxy spectrum is supposed uniform over
its whole extent.  Using this approach, we will investigate the application of
slitless spectroscopy to single object studies to measure internal kinematic
parameters, namely the plateau velocity $v_{0}$ \citep{kalinova_towards_2017,
  varidel_sami_2019} and the central velocity gradient (CVG) $w_{0}$
\citep{lelli_scaling_2013, erroz-ferrer_h_2016}.

The article is organized as follows.  In Sec.~\ref{sec:resolv-kinem-slitl} we
present galaxy kinematics in slitless spectroscopy and describe the model
parametrization.  We test our method on simulated spectrograms in
Sec.~\ref{sec:test_simulations}, and apply it on selected galaxies from 3D-HST
and GLASS survey in Sec.~\ref{sec:data}.  We discuss the results in
Sec.~\ref{sec:discussions}, and conclude and open some perspectives in
Sec.~\ref{sec:conclusion-perspective}.

\section{Resolved kinematics in slitless spectroscopy}
\label{sec:resolv-kinem-slitl}

In this section, we investigate the kinematic signature in a slitless resolved
galaxy spectrum.  The internal velocity induces differential Doppler shifts (in
addition to the systemic cosmological redshift), inducing small offsets of
observed wavelengths as a function of position, and therefore distorting the
overall spectral shape.

The slitless spectrogram $I(x, y)$ can be derived from two key ingredients,
first the spectro-spatial flux distribution cube of the galaxy
$C(\vec{r}, \lambda)$, which contains all the observable information -- spatial
profile, intrinsic spectrum, velocity field, instrumental transmission and PSF,
etc. -- and second the 2D dispersion law $\vec{D}(\lambda)$ from the
spectrograph, relating the wavelength to the $(x, y)$-offset on the detector.
More details on how the spectrogram is computed is given in
Sec.~\ref{sec:galaxy-model}.

\subsection{Pedagogic case}
\label{sec:pedagogic}

To illustrate the effect of resolved kinematics in slitless spectroscopy, we
build a pedagogic simulation mimicking the observation of an \Ha-emitting disk
galaxy at $z \sim 0.9$ with
\begin{itemize}
\item an exponential thin disk density profile, with an inclination of \ang{60}
  and a scale length of $r_{d} = \SI{6}{px}$, and a dispersion direction
  perpendicular to the major axis;
\item a uniform intrinsic spectrum made of a constant continuum and an
    \Ha{} + [\ion{N}{II}] emission line complex at $z = 0.9$;
  \item a typical plateau velocity curve with
    $v \to v_{0} = \SI{300}{km.s^{-1}}$ beyond transition radius
    $r_{0} = \SI{10}{px}$ (see Sec.~\ref{sec:velocity-field}).
\end{itemize}
For illustration purpose, the simulated instrument is similar to an
HST-like slitless spectrograph but with an unrealistic spectral
sampling of $D = \SI{2.5}{\AA.px^{-1}}$, ten times better than actual
slitless instruments.

\begin{figure}
  \centering
  \includegraphics[width=0.9\columnwidth]{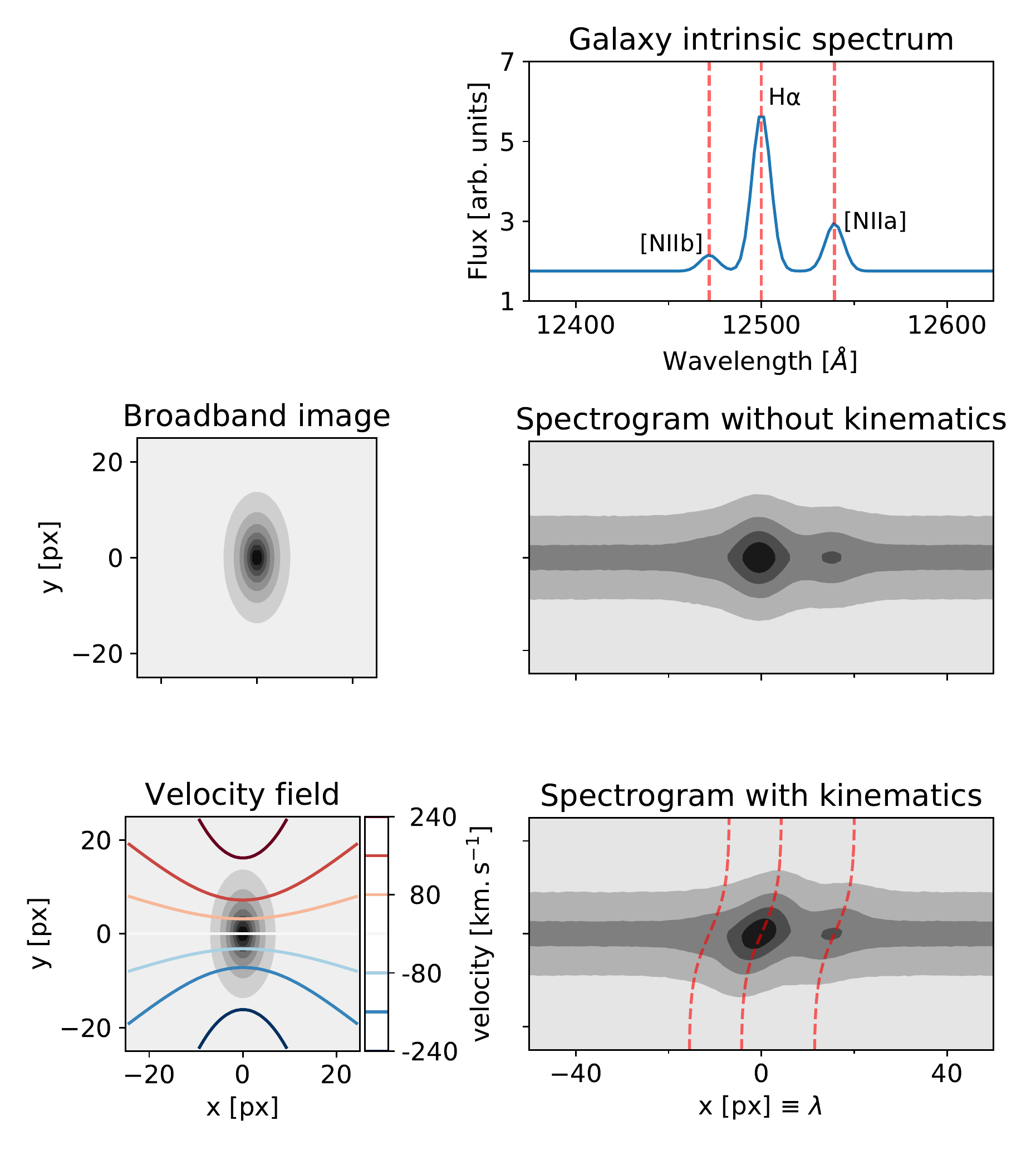}
  \caption{Toy simulation of the spectrogram of a typical
      H$\alpha$-emitting disk galaxy at $z \sim 0.9$ (intrinsic
      spectrum in \emph{uppermost panel}, spatial profile on the
      \emph{leftmost} panels) as observed with an HST-like slitless
      spectrograph but with an improved spectral resolution of
      $\mathcal{R} \sim 2500$ (see text). \emph{Top}: without the
      kinematic effects; \emph{bottom:} the signature of the intrinsic
      velocity field is clearly visible as a distortion of the
      spectrogram. The \emph{red dashed lines} correspond to the
      galaxy Rotation Curves at each emission lines position and are
      tracing the distortion in the spectrogram.}
  \label{fig:slitless_ideal}
\end{figure}

Simulated slitless spectrograms are shown in Fig.~\ref{fig:slitless_ideal}. In
both cases, one can distinguish the \Ha{} + [\ion{N}{II}] emission line complex
from the constant continuum which spreads out on each side. When the velocity
field is ignored (\emph{upper panel}), the emission lines have a shape similar
to the flux distribution; on the opposite, when the kinematic effects are
included (\emph{lower panel}), the velocity-induced Doppler offset
significantly distorts the emission line spectral shape.  As can be seen, the
kinematic signature on the slitless spectrogram is somewhat similar to the one
traditionally observed in long slit spectroscopy.

Two important lessons can be learned from this simple simulation.  In this
particular model, the disc scale length $r_{d}$ has been chosen twice smaller
than the turnover radius of the velocity curve $r_{0}$.  As a consequence, only
the inner solid body rotation part of the velocity field has a clear
observational signature, the plateau region being too far out the exponential
disk extent to have any significant impact on spectrogram shape (see galaxy
rotation curves plotted as \emph{red dashed lines} in the \emph{lower panel} of
Fig.~\ref{fig:slitless_ideal}).  This is discussed more in
Sec.~\ref{sec:test_simulations}.

\begin{figure}
\centering
  \includegraphics[width=0.5\columnwidth]{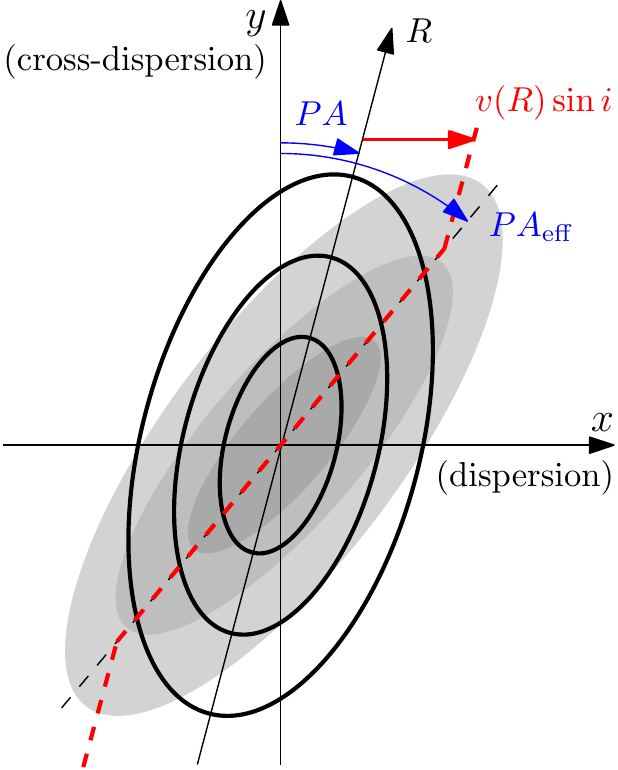}
  \caption{%
      Sketch of the broadband galaxy shape (\emph{open contours}) and the
      spectrogram for an infinitely thin emission line (\emph{shaded contours})
      distorted by intrinsic velocity curve (\emph{red dashed line}); $x$
      (resp.\ $y$) is the direction of dispersion (resp.\ cross-dispersion),
      $PA$ is the intrinsic relative position angle defined as the angle between
      the broadband galaxy major axis and the cross-dispersion direction, and
      $PA_{\mathrm{eff}}$ the apparent position angle on the spectrogram.}
  \label{fig:pa_sketch}
\end{figure}

Secondly, it appears that the \emph{relative} position angle (PA), defined as
the angle between the (projected) galaxy major axis and the cross-dispersion
direction, plays a critical role in the kinematic signature, $PA = 0$ (the
major axis is perpendicular to the dispersion direction) being the most
favorable case.  In the central region of the galaxy where the kinematics is
dominated by the solid body rotation, we can approximate the effective PA
observed in the spectrogram ($PA_{\mathrm{eff}}$) from the relative PA of the
galaxy and the CVG $w_{0}$ (see Fig.~\ref{fig:pa_sketch}):
\begin{equation}
  \label{eq:PA_w0}
  \tan PA_{\mathrm{eff}} \simeq
  \tan PA + \frac{sw_{0}}{R_{kin}} \sqrt{1 + \tan^{2}PA}
\end{equation}
where $s$ is the spatial sampling of the instrument (in \si{arcsec.px^{-1}}),
$w_{0}$ is expressed in \si{km.s^{-1}.arcsec^{-1}}, and kinematic sampling
$R_{kin}$ is defined below in Eq.~\eqref{eq:1}.

Since relative PA and CVG $w_{0}$ are so correlated, they cannot be constrained
independently from the spectrogram alone: the kinematic major axis needs to be
set \emph{a priori} from external photometric observations (more discussions in
Sec.~\ref{sec:velocity-field}).

Overall, due to self confusion effects inherent to slitless spectroscopy, the
kinematic parameters are entangled with spectro-spatial flux distribution of
the galaxy. This is the goal of our analysis to estimate them using an accurate
modeling of the slitless spectrogram.

\subsection{Kinematic sampling}
\label{sec:kinematic-sampling}

The impact of resolved kinematic on slitless spectra can be roughly quantified
using the \emph{kinematic sampling} $R_{kin}$, defined as the line-of-sight
velocity resolution in~\si{km.s^{-1}.px^{-1}}:
\begin{equation}
  \label{eq:1}
  R_{kin} = \frac{D\,c}{\lambda_{o}} \approx \frac{c}{(1+z)\,\mathcal{R}}
\end{equation}
where $D$ is the spectral sampling (in \si{\AA.px^{-1}}),
$\lambda_{o} = (1+z) \lambda_{e}$ the cosmologically redshifted wavelength of a
line at rest-frame wavelength $\lambda_{e}$, and
$\mathcal{R} \equiv \lambda/\Delta\lambda \approx \lambda_{o}/D$ the resolving
power.  As defined, a smaller $R_{kin}$ corresponds to a higher sensitivity to
internal kinematics.

The kinematic sampling estimated for various current and future slitless
surveys is presented in Table~\ref{tab:slitless_table}.  Unfortunately, current
HST-based surveys only have a kinematic sampling
$\gtrsim \SI{800}{km.s^{-1}.px^{-1}}$, which might prove barely sufficient to
derive precise velocity parameters for a large fraction of the sample.  Future
surveys however will all reach
$R_{kin}\sim \SIrange[range-phrase=-]{150}{300}{km.s^{-1}.px^{-1}}$, more
appropriate for detailed kinematic analyses.

\begin{table*}
  \centering
  \caption{Main instrumental properties of past, current and future slitless
    surveys.}
  \label{tab:slitless_table}
  \begin{tabular}{cccccccc}
    \hline \hline
    Survey-telescope & Instrument-grism
    & Spectral range & Spatial scale & $D$ & $\mathcal{R}$ & $R_{kin}$
    & Galaxy number \\
    (1) & (2) & (3) & (4) & (5) & (6) & (7) & (8) \\
    \hline
    GPS$^{a}$-HST & NICMOS(IR)-G141
    & 11000--19000 & 0.20 & 80 & 200 & 1260--2180 & \num{33} \\
    PEARS$^{b}$-HST & ACS-G800L
    & 5500--10500 & 0.05 & 38.5 & 100 & 1100--2100 & \num{10000} \\
    3DHST$^{c}$-HST & WFC3(IR)-G141
    & 10000--17500 & 0.13 & 46.5 & 130 & 800--1400 & \num{100000}\\
    GLASS$^{d}$-HST & WFC3(IR)-G102
    & 7500--11500 &0.13 & 24.5 & 210 & 640--980 & \num{2000} \\
    JWST$^{e}$ & NIRISS (2021)
    & 6000--23000 & 0.065 & 10 & 700 & 130--500 & - \\
    EUCLID$^{f}$ & NISP-R (2022)
    & 12500--18500 & 0.30 & 13 & 380 & 210--310 & \num{30000000} \\
    WFIRST$^{g}$ & GRS (2028)
    & 10000--19500 & 0.11 & 11 & 435-865 & 170--330 & \num{18000000} \\
    \hline
  \end{tabular}
  \tablefoot{(1) Telescope and survey name, (2) instrument and grism, (3)
    wavelength coverage (in \si{\AA}), (4) spatial sampling (in
    \si{arcsec.px^{-1}}), (5) spectral sampling ($=\Delta\lambda$ in
    \si{\AA.px^{-1}}), (6)
    spectroscopic resolving power $\mathcal{R} \equiv \lambda/\Delta\lambda$,
    (7) kinematic sampling for \Ha{} galaxies emitters at spectral range limits
    (in \si{km.s^{-1}.px^{-1}}), (8) approximate catalog size.
    \tablefoottext{a}{HST Grism Parallel survey \cite{thompson_initial_1998}.}
    \tablefoottext{b}{\cite{straughn_emission-line_2008,
        pirzkal_spectrophotometrically_2009}.}
    \tablefoottext{c}{\cite{brammer_3d-hst:_2012, momcheva_3d-hst_2016}.}
    \tablefoottext{d}{\cite{schmidt_through_2014, treu_grism_2015}.}
    \tablefoottext{e}{\cite{doyon_jwst_2012}.}
    \tablefoottext{f}{\cite{grupp_optical_2012}.}
    \tablefoottext{g}{\cite{spergel_wide-field_2015}.} }
\end{table*}

In order to properly constrain galaxy internal kinematics from slitless
spectroscopy, we now build a predictive model sufficiently realistic to be
quantitatively compared to actual observations in a forward approach.  As
noticed earlier, this model only depends on a galaxy model
(Sec.~\ref{sec:galaxy-model}) -- including flux distribution, intrinsic
spectrum and kinematic parameters -- and on an instrumental signature
(Sec.~\ref{sec:instrumental-model}) -- allowing a complete simulation of the
observed spectrogram.

\subsection{Galaxy Model}
\label{sec:galaxy-model}

\subsubsection{Core assumptions}
\label{sec:assumptions}

By design, slitless observations essentially target \Ha{}-emitting
galaxies at high redshift ($0.5<z<1.5$, see Sec.~\ref{sec:data}).
Additionally, given the limited spatial resolution and kinematic
sampling (Table~\ref{tab:slitless_table}), the observations cannot yet
constrain elaborate models.  Therefore, we are making the physical
assumption (yet reasonable for galaxies considered in this analysis)
of an axi-symmetric thin cold disk.  As a consequence, the galaxy flux
distribution, spectrum and velocity field only depends on the internal
radius $R$.

We further adopt the \emph{separability assumption}: the rest-frame galaxy
spectrum is supposed to be uniform over its whole extent, and only modulated by
Doppler shift from internal kinematics.  As a result, the observer-frame galaxy
datacube $C(\vec{r},\lambda)$ -- with two spatial dimensions and a spectral one
-- can easily be reconstructed from the normalized spatial flux distribution
$F(\vec{r})$, the cosmologically redshifted intrinsic spectrum $S(\lambda)$ and
the line-of-sight velocity field $v(\vec{r})$ of the galaxy:
\begin{equation}
  \label{eq:2}
  C(\vec{r}, \lambda) =
  F(\vec{r}) \cdot S\left(\frac{\lambda}{1 + v(\vec{r})/c} \right).
\end{equation}

\subsubsection{Spatial flux distribution}
\label{sec:density-profile}

As mentioned earlier, slitless spectrography is plagued by
self-confusion, mixing different spatial and spectral contributions
from the target on the same part of the detector.  A key component of
our model is therefore a precise description of the internal flux
distribution $F(\vec{r})$ of the resolved galaxy at considered
wavelengths.

Under the assumption of an axi-symmetric thin cold disc, the galaxy
morphology is characterized by its inclination, position angle and
intrinsic radial flux profile.  Such a model will be used to estimate
$F(\vec{r})$ in simulations (Sec.~\ref{sec:test_simulations}).

For real observations (Sec.~\ref{sec:results}), however, given the
morphological variety of galaxies -- e.g.\ spiral arms or disc warps -- we
choose to estimate the galaxy flux distribution directly from the thumbnail
broadband images $B(\vec{r})$ systematically acquired along slitless
spectroscopic observations.

We note that, even though our model will essentially be constrained by
\emph{emission lines} (see Sec.~\ref{sec:spectrum}), the broadband image,
acquired in a band covering the spectrograph band pass, is a direct observation
of the integrated flux mostly originating from the \emph{continuum}.  For high
redshift galaxies ($z \gtrsim 1$), the spatial \Ha{} emission line profile is
shown to be similar to the continuum but more extended on average
\citep{nelson_spatially_2012}.  To allow for this difference, we relate the
internal flux distribution of the galaxy $F(\vec{r})$ to the peak-normalized
broadband image $B(\vec{r})$ through a simple power law:
\begin{equation}
  \label{eq:4}
  F(\vec{r}) = B^{\eta}(\vec{r}),
\end{equation}
where a flux distribution index $\eta < 1$ corresponds to a more diffuse
distribution than broadband one's.

We discuss further the impact of our choices -- in particular the fact that the
flux distribution might not be the same in the continuum and in the emission
line -- in Sec.~\ref{sec:discussions}.

\subsubsection{Intrinsic spectrum}
\label{sec:spectrum}

The 3D-HST and GLASS surveys use grisms G141 and G102 in the infrared domain
and cover the \SIrange[range-units=single]{7500}{17500}{\AA} wavelength range
(see Sec.~\ref{sec:3dhst-glass-survey}).  As shown in
Sec.~\ref{sec:kinematic-sampling}, the kinematic impact is expected to be at
most subtle in the slitless spectrograms, and only significant for strong
emission lines, namely the complex \Ha{}$\lambda 6563$ +
[\ion{N}{II}]$\lambda\lambda 6548, 6584$ +
[\ion{S}{II}]$\lambda\lambda 6718, 6732$ (for a redshift $0.3 < z < 1.7$), or
the doublet [\ion{O}{III}]$\lambda\lambda 4959, 5007$ (for $0.7 < z < 2.5$). We
do not consider fainter emission lines such as H$\beta$.

The intrinsic galaxy spectrum is modeled as a sum of individual
Gaussian lines on top of a smooth continuum:
\begin{equation}
  \label{eq:5}
  S(\lambda) =
  \sum_{\text{line}\;i} \frac{A_{i}}{\sigma\sqrt{2\pi}} \,
  \exp\left(-\frac{(\lambda - (1+z)\lambda_{i})^{2}}{2\,\sigma^{2}}\right) +
  \mathcal{C}(\lambda)
\end{equation}
where $A_{i}$ and $\lambda_{i}$ are respectively the amplitude and the
rest-frame wavelength for each line~$i$, $\sigma$ the supposedly
constant line width, and $\mathcal{C}(\lambda)$ an \emph{ad hoc}
continuum.  For the line doublets, we assume a constant amplitude
ratio of
$A_{[\ion{N}{II}]\lambda 6584} / A_{[\ion{N}{II}]\lambda 6548} =
A_{[\ion{O}{III}]\lambda 5008} / A_{[\ion{O}{III}]\lambda 4960} = 3$
and
$A_{[\ion{S}{II}]\lambda 6718} / A_{[\ion{S}{II}]\lambda 6732} = 1$.

Since the spectrogram adjustment is performed on a very restricted
range around the modeled emission lines, and the instrumental
transmission is assumed to be known (see
Sec.~\ref{sec:dispersion-law}), the continuum $\mathcal{C}$ is simply
modeled by a constant $\mathcal{C}_0$ which is the same for all lines.

Overall, only a handful of parameters are needed to describe the
intrinsic spectrum, namely 6 for an \Ha{}-emitting galaxy: effective
redshift $z$, line amplitudes $A_{\Ha{}}$,
$A_{[\ion{N}{II}]\lambda 6584}$ and $A_{[\ion{S}{II}]\lambda 6718}$,
effective dispersion $\sigma$ and continuum constant $\mathcal{C}_0$.

\subsubsection{Velocity field}
\label{sec:velocity-field}

In this section, we present how we construct a model for the galaxy velocity
field $v(\vec{r})$.  Even under simplifying hypotheses, a physical modelling of
galaxy rotation curves (hereafter RC) requires a detailed description of the
contributions from the disc, bulge and halo components to the galaxy dynamics,
to be constrained only with high precision morphologic and spectroscopic
observations \citep[e.g.][]{courteau_optical_1997}, out of reach to
low-dispersion slitless spectrography.

Under the assumption of an axi-symmetric thin cold disc, one can revert to an
analytic expansion to reproduce the overall shape of the intrinsic rotation
velocity curve $v_{\text{rot}}(r)$ with a restricted number of empirical
parameters. In our case, we use a simple hyperbolic tangent profile, very
similar to the commonly used arc-tangent profile \citep{stott_kmos_2016,
  pelliccia_hr-cosmos:_2017}:
\begin{align}
  \label{eq:7}
  v_{\text{rot}}(r) &= v_{0} \, \tanh \, \left(\frac{r}{r_{0}} \right) \\
  &= v_{0} \, \tanh \, \left(\frac{w_{0} \, r}{v_{0}} \right)
    \approx
    \begin{cases}
        w_{0}\,r & r \ll r_{0} \quad\text{(solid body)} \\
        v_{0} & r \gg r_{0}  \quad\text{(plateau)}
    \end{cases}
    \label{eq:w0v0}
\end{align}
where $v_{0}$ is the plateau value of the RC and $r_{0}$ is the transition (or
turnover) radius.  In practice, we use expression~(\ref{eq:w0v0}), since the
CVG $w_{0} \equiv v_{0}/r_{0}$ is the dominant term at small radius, leading to
less degeneracy between the parameters.

We note that $v_{0}$ can only be significantly constrained if the
observations go beyond $r \sim r_{0}$.  As briefly explained in
Sec.~\ref{sec:resolv-kinem-slitl}, there is a competition between the
RC turnover radius $r_{0}$ and the galaxy disc scale length $r_{d}$:
if $r_{0} \lesssim r_{d}$, then both $v_{0}$ and $w_{0}$ can be
reasonably constrained; alternatively, if $r_{0}$ is significantly
larger than $r_{d}$, only the solid body rotation parameter $w_{0}$
can be sensibly measured.

For a thin cold disk, the observed mean velocity field $v(\vec{r})$ along the
line of sight is straightforwardly given by: 
\begin{equation}
  \label{eq:10}
  v(\vec{r}) = v(x,y) = cz + v_{\text{rot}}(R)\cos\theta \sin i
\end{equation}
where $cz$ is the systemic velocity, $v_{\text{rot}}(R)$ is the rotation curve,
$i$ the galaxy inclination and $\theta$ the azimuthal angle in the plane of
the galaxy:
\begin{align}
  R \cos i \sin\theta &= - (x - x_{0})\cos(PA) - (y - y_{0})\sin(PA) \\
  R \cos\theta &= -(x - x_{0})\sin(PA) + (y - y_{0})\cos(PA)
  \label{eq:11}
\end{align}
with PA the relative position angle, $\vec{r} = (x, y)$ the Cartesian
coordinates in the sky, and $(x_{0}, y_{0})$ the galactic center
coordinates. The galaxy being modeled as a cold rotating thin disc, the
kinematic and morphologic position angles are assumed to be the same.

Since PA is highly degenerate with $w_{0}$ (see Fig.~\ref{fig:slitless_ideal}),
it is crucial to constrain it independently from photometry.  To do so, we
estimate the projection angle $i$ and relative position angle PA from a Sersic
fit \citep{sersic_atlas_1968} to the broadband image used for the flux
distribution.  However, since only $v_{\text{rot}}(R)\,\sin i$ is adjusted, we
stress out that the inclinaison $i$ is not needed in the fit \emph{per se}, but
only for post-analysis velocity deprojection if needed.

\subsection{Instrumental model}
\label{sec:instrumental-model}

The second key ingredient needed to simulate the slitless spectrogram is the
dispersion law $\vec{D}(\lambda)$ from the spectrograph, as well as its
transmission curve $\mathcal{T}(\lambda)$.

\paragraph{Dispersion law.}
\label{sec:dispersion-law}
The dispersion law $\vec{D}(\lambda)$ gives the $(x, y)$-offset on the detector
(with respect to a reference position) as a function of wavelength.  This is
mostly an instrumental quantity, derived from dedicated calibration
procedures \citep[e.g.][]{kuntschner_wfc3_2009, kuntschner_wfc3_2009-1}.

Even though the forward approach described in Copin (\emph{in prep.}) would be
an appropriate way to calibrate the dispersion law, we rely in this analysis on
the WCS solution computed and delivered for each galaxy by standard data
reduction.  Given the required precision of our model, we observed minor
inconsistencies for some spectrograms, in the form of a px-scale offset in the
cross-dispersion direction, as a result of a small registration error between
broadband and dispersed images.  To account for this effect, we introduce a
nuisance parameter $\Delta y$ (in~px). We note that a similar mis-registration
along the dispersion axis is corrected to first-order by the effective
redshift~$z$ adjusted in the procedure.

\paragraph{Transmission.}
\label{sec:transmission}
The transmission $\mathcal{T}(\lambda)$ conveys the chromatic evolution of the
instrumental response, and is derived from specific flux calibration
\citep{kuntschner_revised_2011}.  In our model, we simply include the provided
transmission into the galaxy datacube $C(\vec{r}, \lambda)$ derived from
Sec.~\ref{sec:galaxy-model}.

\subsection{Forward modeling}
\label{sec:forward-modeling}

\subsubsection{Spectrogram reconstruction}
\label{sec:model-spectrogram}

Once the galaxy datacube $C(\vec{r}, \lambda)$ and the dispersion law
$\vec{D}(\lambda)$ are known, we compute the resulting spectrogram using 
\citep[e.g.][]{freudling_hubble_2008}:
\begin{equation}
  I(\vec{r}) = \int\d\lambda\,C(\vec{r} - \vec{D}(\lambda), \lambda)
\end{equation}
In practice, the spatial convolution and wavelength integration are
performed in Fourier space (Copin, \emph{in prep.}).

Since both quantities $C(\vec{r}, \lambda)$ and $\vec{D}(\lambda)$ depends on
various parameters $\vec{p}$, we are able to simulate a spectrogram to be
compared to the observations.  This ``forward'' approach allows to constrain
physical or instrumental parameters directly in the observation space, in
opposition to more traditional ``inverse'' methods extracting fully or
partially free parameters from data using \emph{ad hoc} procedures.

\subsubsection{Maximum likelihood}
\label{sec:adjust-procedure}

For an observed spectrogram $D_{ij}$ with estimated variance
$\sigma^{2}_{ij}$, we compute a spectrogram model $I_{ij}(\vec{p})$
and constrain the set of free parameters $\vec{p}$ by minimizing the
following $\chi^{2}$:
\begin{equation}
    \label{eq:14}
    \chi^{2}(\vec{p}) = \sum_{\text{px}\;ij}
    \left(\frac{D_{ij} - I_{ij}(\vec{p})}{\sigma_{ij}}\right)^{2}.
\end{equation}
The maximum likelihood procedure further provides the full parameter covariance
matrix, from which one can compute the $1\sigma$ uncertainties.  Since only the
emission lines are significantly distorted by resolved kinematics, the
adjustment is performed on a restricted area about $40 \times \SI{60}{px}$
around emission lines of interest, equivalent to
$\ang{;;5.2} \times \SI{700}{\AA}$ (resp. $\SI{1400}{\AA}$) for galaxy
respectively observed with grism G102 (resp. G141).

\section{Validation on simulations}
\label{sec:test_simulations}

\subsection{Fiducial case}
\label{sec:fiducial-case}

We present realistic simulations designed to mimic WFC3-GLASS
observations with grism G102 (see Table~\ref{tab:slitless_table}) in a
favorable case where the disc scale length $r_{d}$ is similar to the
RC turnover radius~$r_{0}$.

\begin{figure}
    \includegraphics[width=\columnwidth]{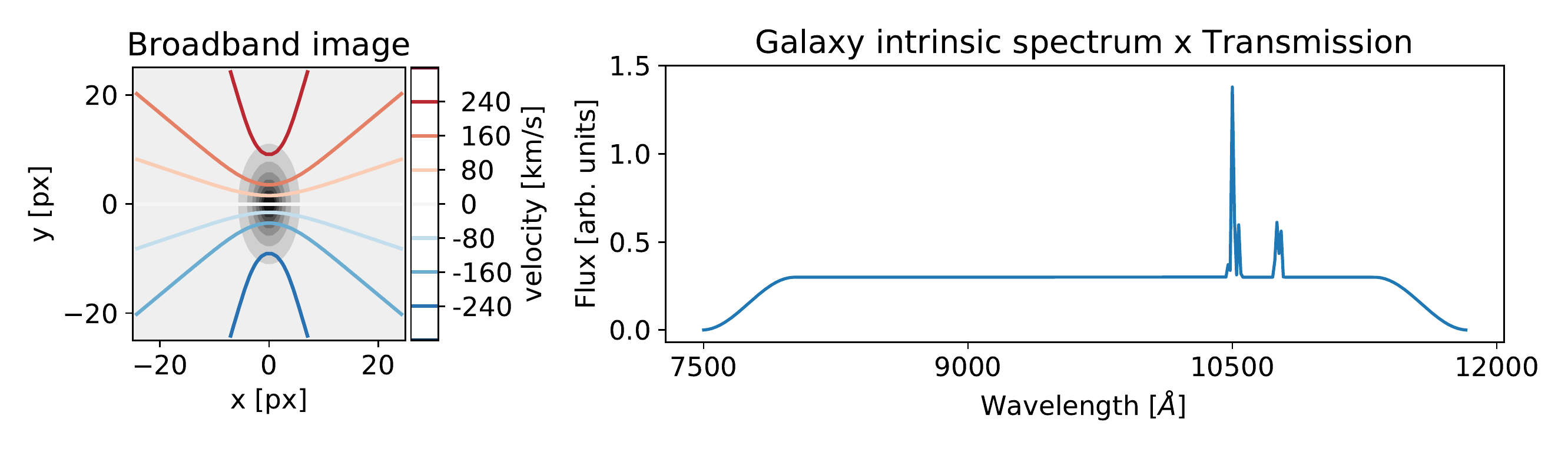}
    \caption{Galaxy model to construct the fiducial test spectrogram.
      \emph{Left}: spatial flux distribution $F(x,y)$ and velocity field
      $v(x,y)$ with $v_{0}\sin i = \SI{250}{km.s^{-1}}$ and
      $w_{0}\sin i= \SI{420}{km.s^{-1}.arcsec^{-1}}$.  \emph{Right}: effective
      galaxy spectrum $S(\lambda)\times \mathcal{T}(\lambda)$, including
      instrumental transmission.}
    \label{fig:simu_model}
\end{figure}

We first simulate spectrograms with the following properties (see
Fig.~\ref{fig:simu_model}):
\begin{description}
\item[flux distribution:] axisymetric exponential profile with a disc scale
  length $r_{d} = \ang{;;0.6} \sim \SI{5}{px}$, an inclination $i = \ang{60}$
  and a relative position angle $PA = \ang{0}$.
\item[intrinsic spectrum:] from
    \SIrange[range-units=single]{7500}{11500}{\AA} with 5~emission
    lines to simulate \Ha{}+[\ion{N}{II}]+[\ion{S}{II}] complex at
    $z=0.6$, with a constant line width $\sigma=\SI{5}{\AA}$ and a
    constant continuum;
\item[velocity field:] $v_{0} \sin i = \SI{250}{km.s^{-1}}$ and
    $w_{0}\sin i = \SI{420}{km.s^{-1}.arcsec^{-1}}$ corresponding to a
    turnover radius $r_{0} = r_{d}$;
\item[instrumental model:] linear dispersion law $\vec{D}(\lambda)$
    aligned along the $x$-axis, and simplified apodized transmission.
\end{description}
A constant gaussian noise component is finally added, so that the Peak
Signal-to-Noise Ratio (PSNR) is around 40 (a typical value for high
quality slitless HST spectra).

\begin{figure}
  \includegraphics[width=\columnwidth]{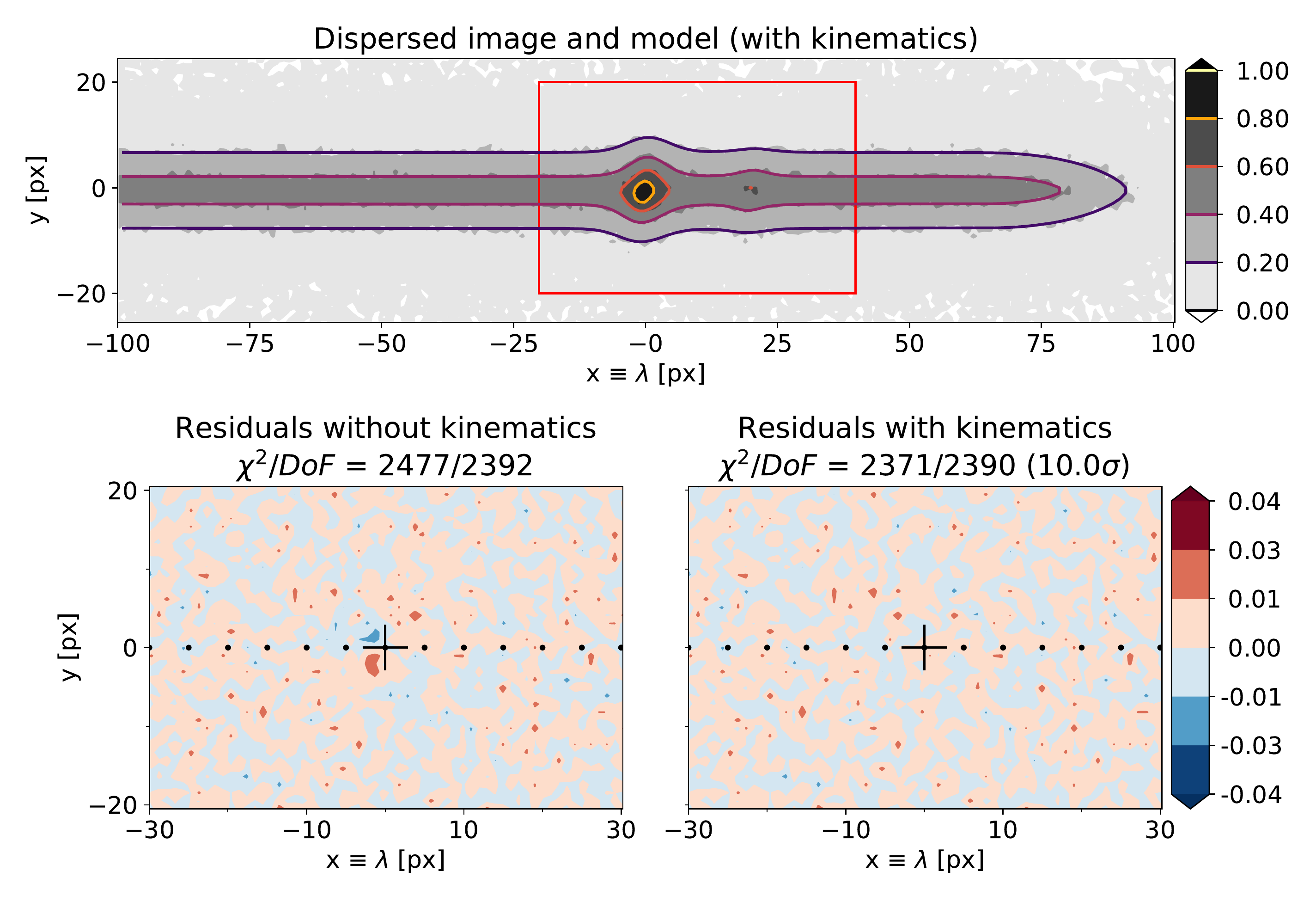}
  \caption{%
    \emph{Top:} input peak-normalized dispersed image data (\emph{gray}) and
    the model with kinematics (\emph{contours}) adjusted on the fit region
    (\emph{rectangle}). \emph{Bottom left:} residual map for the model without
    kinematics, with a quadrupolar structure visible at \Ha{} line
    location. \emph{Bottom right:} residual map for the model including
    kinematics. The \emph{cross} represents the position and relative PA of the
    galaxy at \Ha{} wavelength. The decrease in the $\chi^{2}$ with two
    additional kinematic parameters corresponds to a $10.0\sigma$~detection.}
  \label{fig:model_residu}
\end{figure}

The fit residuals without (resp.\ with) kinematics is shown in the middle
(resp.\ bottom) panel of Fig.~\ref{fig:model_residu}.  The minimal $\chi^{2}$
computed on a rectangular region of $40 \times 60$~px around the \Ha{} and
[\ion{S}{II}] lines decreases from 2688 for $40 \times 60 - 7 = 2392$~degrees
of freedom (DoF) to 2380 with only two additional kinematic parameters
$(w_{0}, v_{0})$.  This $\Delta\chi^{2} = -106$ decrement for a model with only
two extra parameters has a one-tailed $p$-value of $\sim \num{e-23}$,
corresponding to a $10.0\sigma$ detection level.  As expected, the residual map
without kinematics displays a clear signature of the velocity field as a
coherent quadrupolar structure around \Ha{} line position; this structure
disappears in the residual map with kinematics.

The fiducial fit with kinematics gives
$v_{0}\sin i = \SI{243 \pm 39}{km.s^{-1}}$,
$w_{0}\sin i = \SI{397 \pm 53}{km.s^{-1}.arcsec^{-1}}$ and
$z = \num{0.59998 \pm 0.00004}$ (statistical error), all within
$1\sigma$ of the input values.

\begin{figure}
  \includegraphics[width=\columnwidth]{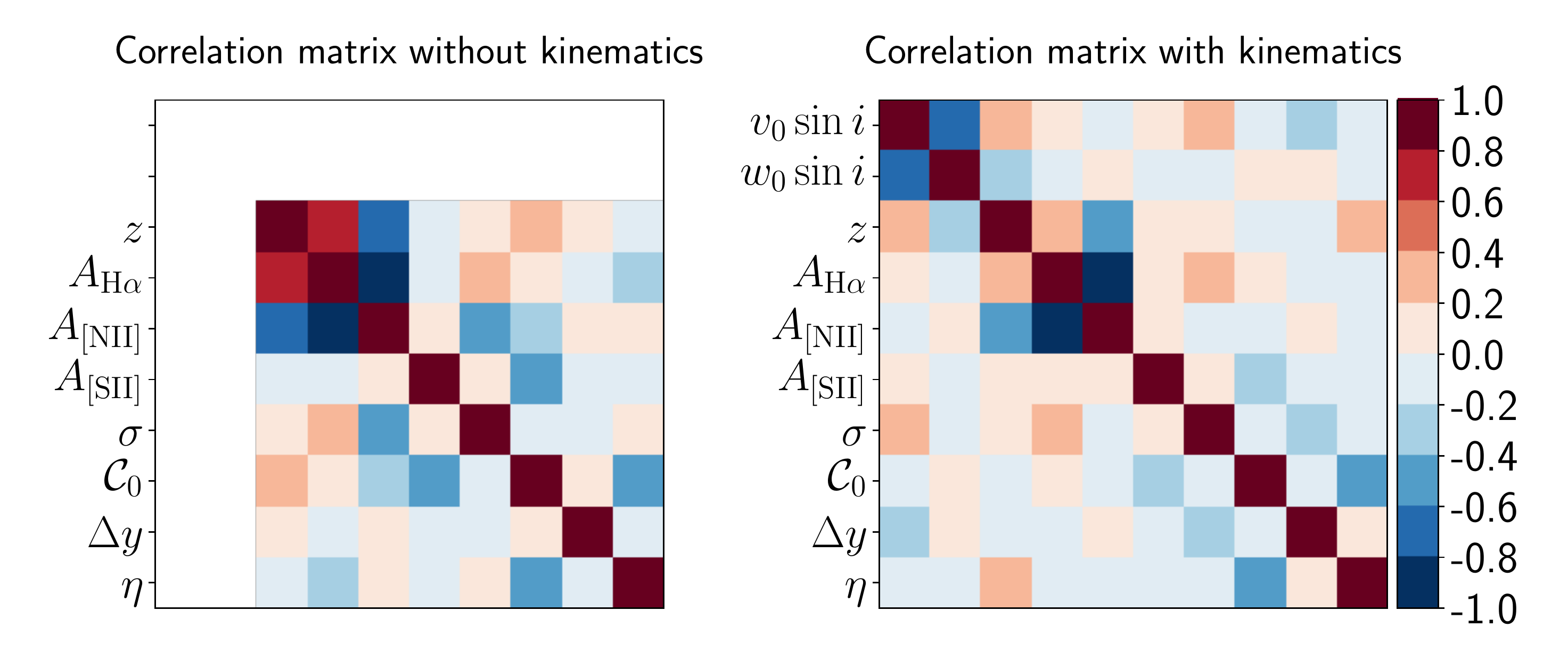}
  \caption{Correlation matrix for both fits without (\emph{left}) and with
    (\emph{right}) kinematics for the fiducial simulation.}
  \label{fig:correlations_simu}
\end{figure}

The correlation matrices are presented in Fig.~\ref{fig:correlations_simu} for
the fit without and with kinematics.  Note that the kinematics parameters
$v_{0}\sin i$ and $w_{0}\sin i$ are slightly anti-correlated, but are almost
uncorrelated to the other adjusted parameters: \emph{no other galactic or
  instrumental parameter can mimic a kinematic signature}.

To test the parameter distribution, we perform the fit for 500 different
realizations of the gaussian noise with a PSNR of 40 in the same configuration,
and we present the marginalized distributions of measured parameters
$v_{0}\sin i$, $w_{0}\sin i$ and redshift $z$ in Fig.~\ref{fig:model_residu}.
All distributions are consistent with the input parameters in the simulation,
and reasonably gaussian.  As expected from correlation matrix
(Fig.~\ref{fig:correlations_simu}), there is a slight anti-correlation between
kinematic parameters.

\begin{figure}
  \includegraphics[width=\columnwidth]{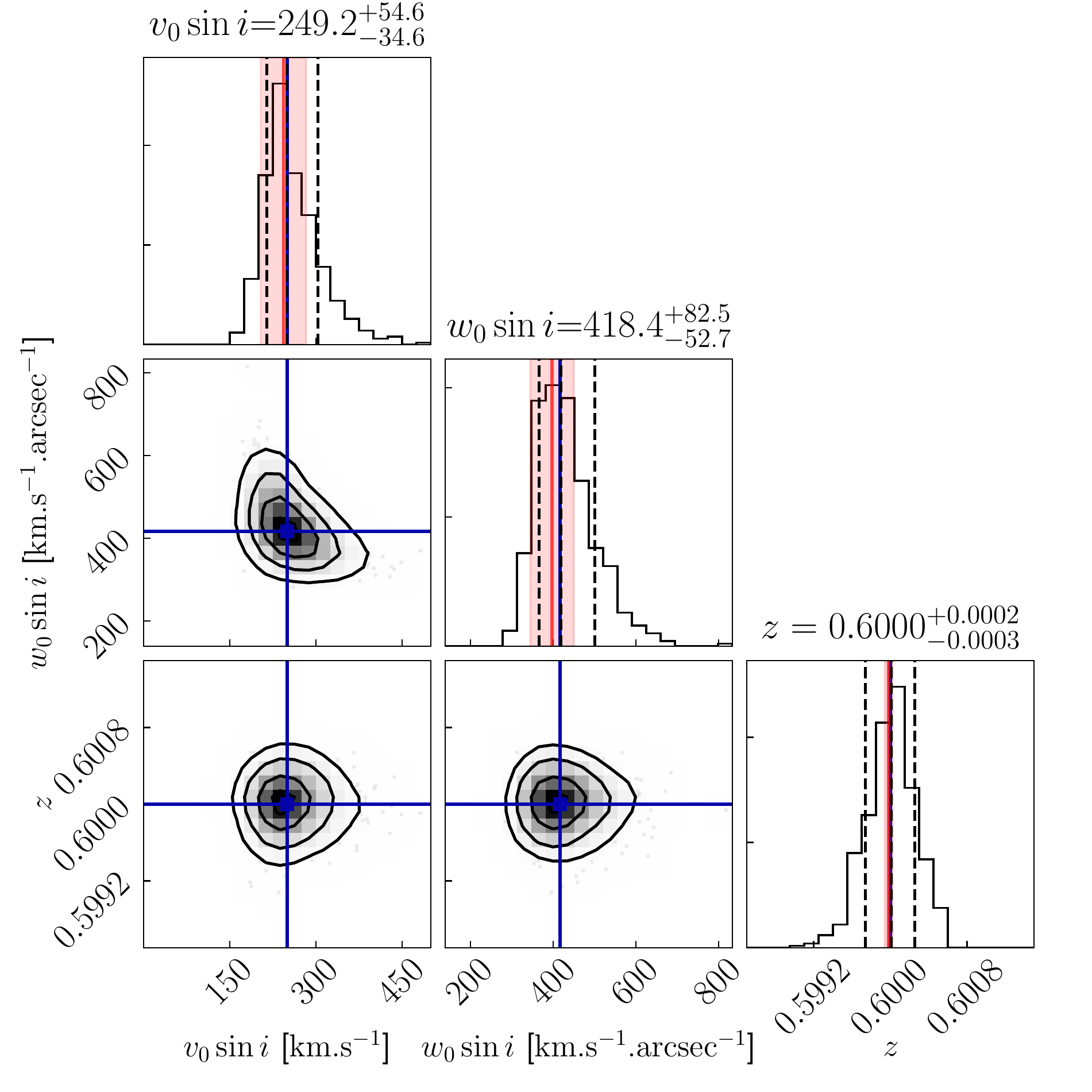}
  \caption{%
    Marginalized distributions of the kinematic parameters $v_{0}\sin i$,
    $w_{0}\sin i$ and redshift $z$ for 500 realizations of a gaussian noise
    with PSNR of 40.  The \emph{blue lines} show the input values for the
    simulation, the \emph{dashed lines} show the $16^{\text{th}}$,
    $50^{\text{th}}$ and $84^{\text{th}}$ percentiles of the posterior
    distribution. The \emph{red lines} (resp. \emph{shadded
      region}) indicate the fit result (resp. $\pm1\sigma$ error band) derived
    from the fiducial noise realization.}
  \label{fig:cornerplot_fiducial_psnr40}
\end{figure}

\subsection{Impact of simulation parameters}
\label{sec:impact-parameters}

The following paragraphs are dedicated to different effects which can impact
the measure of both the kinematic parameters and redshift.

\paragraph{Impact of Signal-to-Noise Ratio.}
We consider the same model as described previously, but perform the fit
for 500 realizations of a gaussian noise with a lower PSNR of
20. As expected, the kinematic parameter marginalized distributions are
broader around the true values, but there is no hint of biases at lower SNR.

\begin{figure*}
  \includegraphics[width=.5\textwidth]{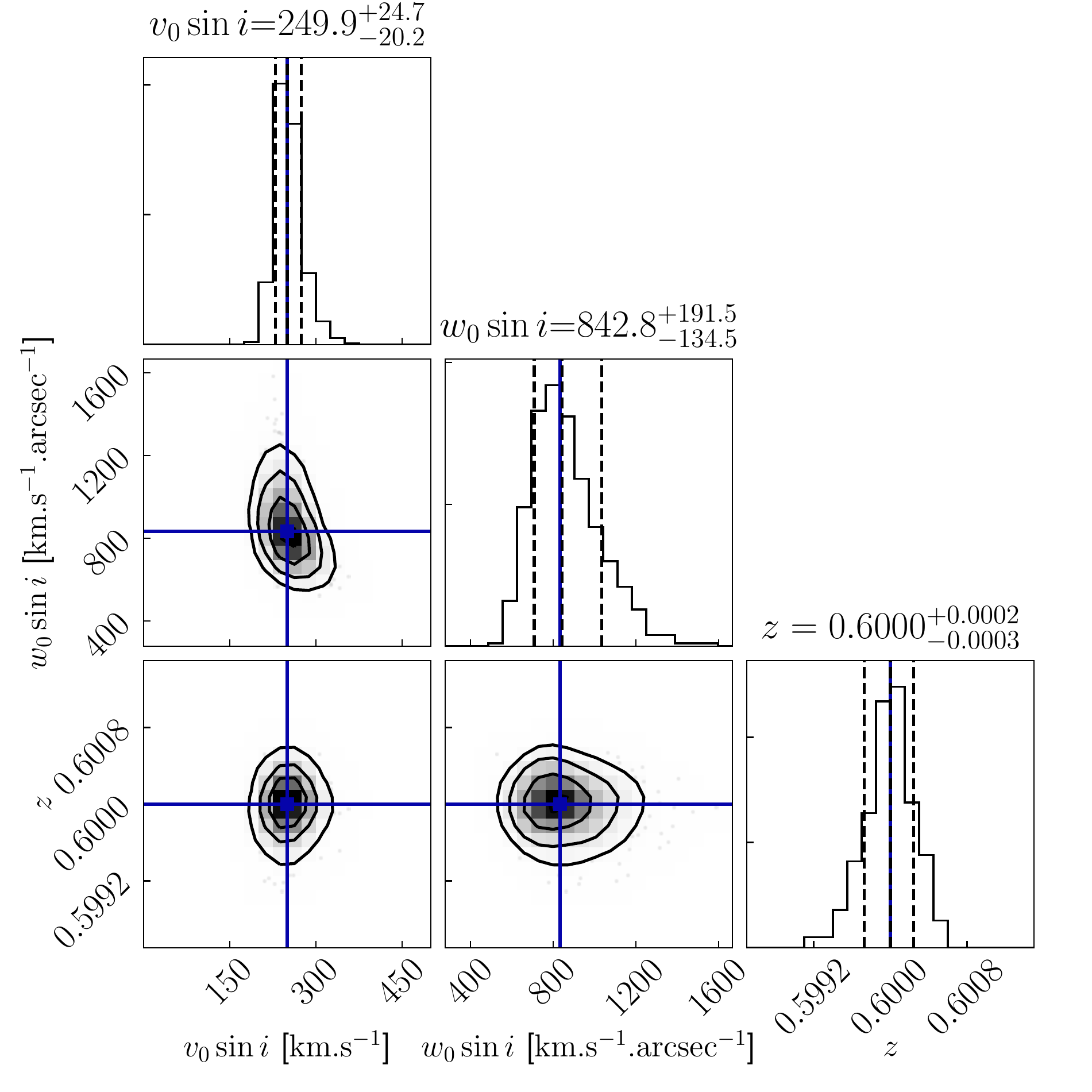}
  \includegraphics[width=.5\textwidth]{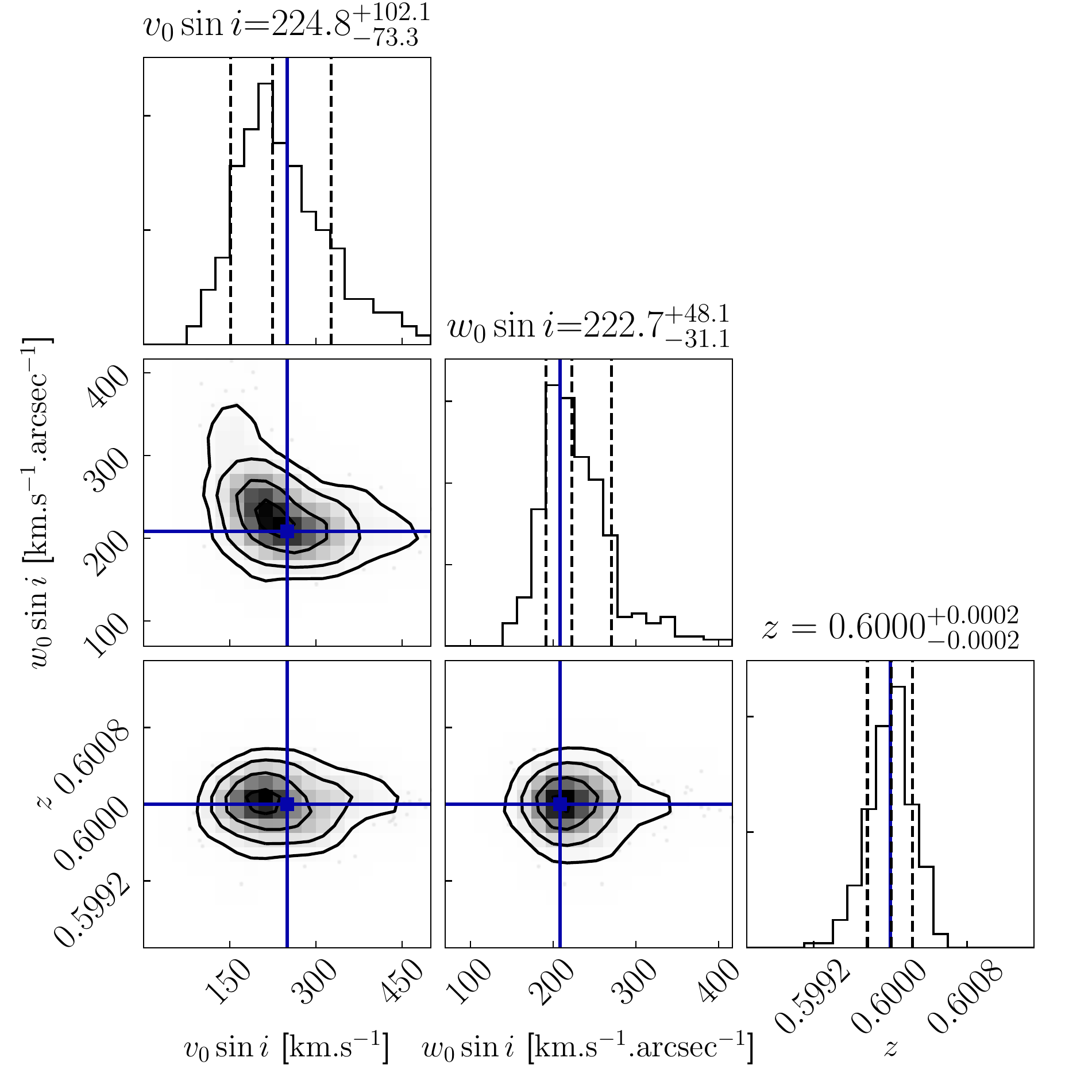}
  \caption{Same as Fig.~\ref{fig:cornerplot_fiducial_psnr40} with
    $r_{0} = 1/2\,r_{d}$ (\emph{left},
    $w_{0}\sin i = \SI{840}{km.s^{-1}.arcsec^{-1}}$) or $r_{0} = 2\,r_{d}$
    (\emph{right}, $w_{0}\sin i =\SI{210}{km.s^{-1}.arcsec^{-1}}$). As noticed
    earlier, the plateau velocity $v_{0}$ is barely constrained when
    $r_{d} \lesssim r_{0}$ (\emph{right}), i.e.\ when the disk extent only
    covers the inner solid-rotation part of the velocity field.  This has
    however only a little impact on the determination of redshift~$z$ and
    CVG~$w_{0}$.}
  \label{fig:cornerplot_rd_r0}
\end{figure*}

\paragraph{Impact of $r_d/r_0$.}
We construct simulations with the same attributes as the fiducial case but with
$r_{d}/r_{0}= 1/2$ or $2$, with fixed $r_{d} = \ang{;;0.6}$ and
$v_{0}\sin i = \SI{250}{km.s^{-1}}$.  The marginalized distributions of
$v_{0}\sin i$, $w_{0}\sin i$ and redshift $z$ are presented in
Fig.~\ref{fig:cornerplot_rd_r0}.  As previously illustrated in the pedagogic
case (Sec.~\ref{sec:pedagogic}), the case $r_{0}/r_{d} = 1/2$ (\emph{left
  panels}) is favorable: the plateau of the galaxy RC is reached within the
photometric extent of the galaxy, and the maximum velocity $v_{0} \sin i$ is
well measured.  In the opposite case $r_{0}/r_{d} = 2$ (\emph{right panels}),
the velocity turn-over radius lays outside the extent of the galaxy disc, and
velocity $v_{0} \sin i$ is only marginally constrained.  However, since the
inner solid body rotation part always lies within the disc extent, the CVG
$w_{0} \sin i$ is always estimated with a similar accuracy.

\section{Application to HST-based observations}
\label{sec:data}

In this section, we present the results obtained by applying the forward model
to real galaxy spectrograms from both 3D-HST and GLASS surveys.

\subsection{3D-HST and GLASS surveys}
\label{sec:3dhst-glass-survey}

The 3D-HST survey is a 248-orbit HST Treasury program to measure
Wide-Field-Camera~3 (WFC3) G141 grism spectra in four of the five deep fields of
the CANDELS Multi-Cycle Treasury project \citep[AEGIS, COSMOS, GOODS-S and
UDS,][]{grogin_candels:_2011, koekemoer_candels:_2011} conducted between 2010
and 2012 \citep{brammer_3d-hst:_2012}. The WFC3 G141 grism has a spectral
coverage from \SIrange{1.0}{1.75}{\um} (corresponding to ground-based $J$ and
$H$ bands) with a dispersion about \SI{46}{\AA.px^{-1}} (twice smaller in the
actual spectrograms after resampling), corresponding to a spectral resolution of
$\sim 130$ in the primary spectral order. This survey also observed in parallel
mode with the instrument ACS using the grism G800L covering
\SIrange{0.5}{1.0}{\um} wavelengths. More details on this survey, observation
strategy and data reduction are given in \citep{van_dokkum_first_2011,
    skelton_3d-hst_2014, momcheva_3d-hst_2016}.

The GLASS survey is a HST large program (140 orbits) with the goal of measuring
grism spectra over the field of ten massive galaxy clusters at redshift
$z = 0.31-0.69$ (for H$\alpha$ emitters) \citep{schmidt_through_2014,
  treu_grism_2015}. The WFC3 is also used in this survey covering wavelength
between 0.75 and \SI{1.75}{\um} using both grism G102 and G141 to observe the
cluster cores. The spectra are acquired at two
almost orthogonal roll angles $\ang{90 \pm 10}$ to ease
cross-decontamination. It results in a catalog of 1272 redshifts down to
$M_{AB} \leq 26$ (1060 redshifts with $M_{AB} \leq 24$). The WFC3 G102 grism
has a spectral coverage from \SIrange{0.75}{1.15}{\um}, a dispersion about
\SI{24.5}{\AA.px^{-1}} (twice smaller in resampled spectrograms) and a
resolution of 210 in the primary spectral order. The analysis was
performed using the 3D-HST reduction pipeline \citep{brammer_3d-hst:_2012}.

We would have liked to test other HST-based surveys such as FIGS
\citep{pirzkal_figsfaint_2017}, but the reduced spectrograms are not yet
publicly available.  The on-going FIGS survey is providing more than
\num{10000} spectrograms of \num{2000} different sources (each object is
observed at five different roll angles).  It has a huge potential for our
analysis since multi-roll joint adjustments (as presented in
Sec.~\ref{sec:glass-result}) would improve constraints on the redshift
and kinematic parameters.


\subsection{Kinematic sample preselection}
\label{sec:selection-data}

Keeping in mind that the kinematic signature is expected to often be evasive on
the HST spectrograms, we do not foresee to detect it for all targets.  We
therefore applied some criteria on both
3D-HST\footnote{\url{https://3dhst.research.yale.edu/Data.php}} and
GLASS\footnote{\url{https://archive.stsci.edu/prepds/glass/}} catalogs to
select the most promising targets.

From the \num{33559}~initial \Ha{} and [\ion{O}{III}] emitters in the
G102 and G141 wavelength domains, we first selected highly significant
emission lines ($F_{line} > \SI{25e-17}{erg.s^{-1}.cm^{-2}}$ and $>15\sigma$
detection level on flux for 3D-HST and quality factor $Q \geq 3$ for
GLASS). We further selected well-resolved and bright galaxies --
allowing an accurate measurement of relative PA -- by applying
cuts on their effective radius ($r_{e} > \ang{;;0.6} \sim \SI{5}{px}$ at WFC3
scale) and integrated magnitude ($M_{F140W} < 22$), as well as
moderately inclined galaxies ($\ang{20} < i < \ang{80}$): face-on
galaxies have a vanishing apparent velocity field, while edge-on
galaxies are generally not well approximated by a thin cold disk.

A final visual inspection of the spectrograms of the
386~pre-selected candidates was performed to discard severely
contaminated spectra, and other data issues. Broadband images were also
examined to remove galaxies with highly asymmetric flux distributions,
on-going mergers or any other complex structures that could not be
handled within our model assumptions. 

This selection process picked out 87~galaxies (57~with \Ha{} emission line only,
11~with [\ion{O}{III}] only, and 19~with both \Ha{} + [\ion{O}{III}] lines) from
3D-HST survey, and 28~galaxies (24~\Ha{}, 2~[\ion{O}{III}] and 2~\Ha{} +
[\ion{O}{III}]) from GLASS survey.  This sub-sample is only a minimal
pre-selection, from which we present the most promising cases in terms of
kinematic signature. In this proof-of-concept analysis, we do not try
to estimate the overall fraction of targets over which the kinematic
analysis is prone to provide accurate measurements; this should be the
subject of a forthcoming study once the methodology is applied in a
systematic way over large simulated and/or observed samples.

\subsection{Results}
\label{sec:results}

\subsubsection{Fitting procedure}
\label{sec:fitting-method}

The adopted method to probe kinematics from the spectrograms is the
following.
\begin{enumerate}
\item The position angle is estimated from a Sersic fit to the broadband image,
  and is kept fixed afterwards. This preliminary measurement was
  necessary since it is a critical parameter of the model.
\item A kinematic-less fit ($v_{0} = w_{0} \equiv 0$) is performed as a
  reference, providing an estimate of spectral and nuisance parameters.
\item The final fit including kinematic parameters $v_{0}$ and $w_{0}$ is
  performed using previous estimates as initial guess.
\item The one-tailed $p$-value is computed from the best-fit $\chi^{2}$ without
  and with kinematics to assess the significance of the fit improvement with
  the addition of two kinematic parameters. The $p$-value is converted into a
  kinematic detection $z$-score, expressed in~$\sigma$.
\end{enumerate}

In the following sections, we present some particular kinematic
detection by computing the objective $\chi^{2}$ on a rectangular
region of $40 \times 60$~px around the emission line of interest,
\Ha{} complex or [\ion{O}{III}] doublet.
Table~\ref{tab:redshift-fit-precision} summarizes the main parameters
values and uncertainties of the various fits detailed below.

\begin{table*}
  \centering
  \caption{Adjusted parameters for test galaxies from the GLASS and 3D-HST
    surveys.}
  \label{tab:redshift-fit-precision}
  \begin{tabular}{c|S|S|S|S}
    \hline
    Galaxy
    & {GLASS \#1134} & {GLASS \#451} & {GLASS \#399} & {3D-HST \#19843} \\
    Cluster / Field
    & {{MACS1423}} & {{MACS2129}}
    & {{MACS0717}} & {{AEGIS}} \\
    $M_{F140W}$
    & {20.8} & {21.7} & {20.8} & {21.4} \\
    \hline
    Grism
    & {G102} & {G141} & {G141} & {G141} \\
    Emission lines
    & {\Ha{}} & {[\ion{O}{III}]+\Ha{}} & {[\ion{O}{III}] $\times 2$~rolls} & {\Ha{}} \\

    $R_{kin}$ (\si{km.s^{-1}.px^{-1}})
    & 700 & 910 & 1000 & 1100 \\
    \hline
    \hline
    PA (1)
    & 4.6 \pm 0.4 & 154.1 \pm 0.9 & 52.2 \pm 0.4 & 10.4 \pm 0.7 \\
    $z_{\text{HST}}$ (2)
    & 0.55000 \pm 0.00465 & 1.36750 \pm 0.00712
    & 1.68700 \pm 0.00806 & 0.95130 \pm 0.00585 \\
    $z_{\text{no-kin}}$ (3)
    & 0.55237 \pm 0.00016 & 1.36266 \pm 0.00047
    & 1.69088 \pm 0.00025 & 0.95459 \pm 0.00016 \\
    $z_{\text{kin}}$ (4)
    & 0.55201 \pm 0.00002 & 1.36415 \pm 0.00005
    & 1.69175 \pm 0.00006 & 0.95544 \pm 0.00013 \\
    $v_{0}\sin i$ (5)
    & 205 \pm 28 & 360 \pm 25 & 287 \pm 15 & 330 \pm 41 \\
    $w_{0}\sin i$ (6)
    & 242 \pm 28 & 350 \pm 31 & 605 \pm 32 & 970 \pm 271 \\
    $A_{\Ha{}}$ (7)
    & 55.7 \pm 2.9 & 45.9 \pm 4.4 & {--} & 50.7 \pm 1.3 \\
    $A_{\ion{O}{III}}$ (7)
    & {--} & 43.5 \pm 3.0 & 96.9 \pm 1.5 & {--} \\
    $\sigma$ (8)
    & 2.96 \pm 0.15 & 5.69 \pm 0.27
    & 6.38 \pm 0.09 & 9.46 \pm 0.49 \\
    $\eta$ (9)
    & 1.014 \pm 0.012 & 0.981 \pm 0.011
    & 1.048 \pm 0.005 & 1.008 \pm 0.015 \\
    $z$-score (10)
    & 12.7 & 13.2 & 25.5 & 6.2 \\
    \hline
  \end{tabular}
  \tablefoot{%
    (1) Relative position angle in degrees; (2) The
    mean redshift total uncertainty (stat. + syst.) for 3D-HST and GLASS
    surveys is $\sigma_{z} \approx 0.003 \times (1+z)$
    \citep{momcheva_3d-hst_2016}. From here, all parameters are derived from
    the model with kinematics (except $z_{\text{no-kin}}$) and uncertainties
    are only statistical; (3) redshift derived from model without kinematics;
    (4) redshift derived from model with kinematics; (5) plateau velocity in
    \SI{}{km.s^{-1}}; (6) central velocity gradient in
    \SI{}{km.s^{-1}.arcsec^{-1}} (7) emission line flux in
    \SI{e-17}{erg.s^{-1}.cm^{-2}}; (8) emission line width in \SI{}{\AA}; (9)
    spatial flux distribution index; (10) kinematic detection $z$-score in
    $\sigma$.%
  }
\end{table*}

\subsubsection{GLASS results}
\label{sec:glass-result}

\begin{figure*}
  \includegraphics[width=\textwidth]{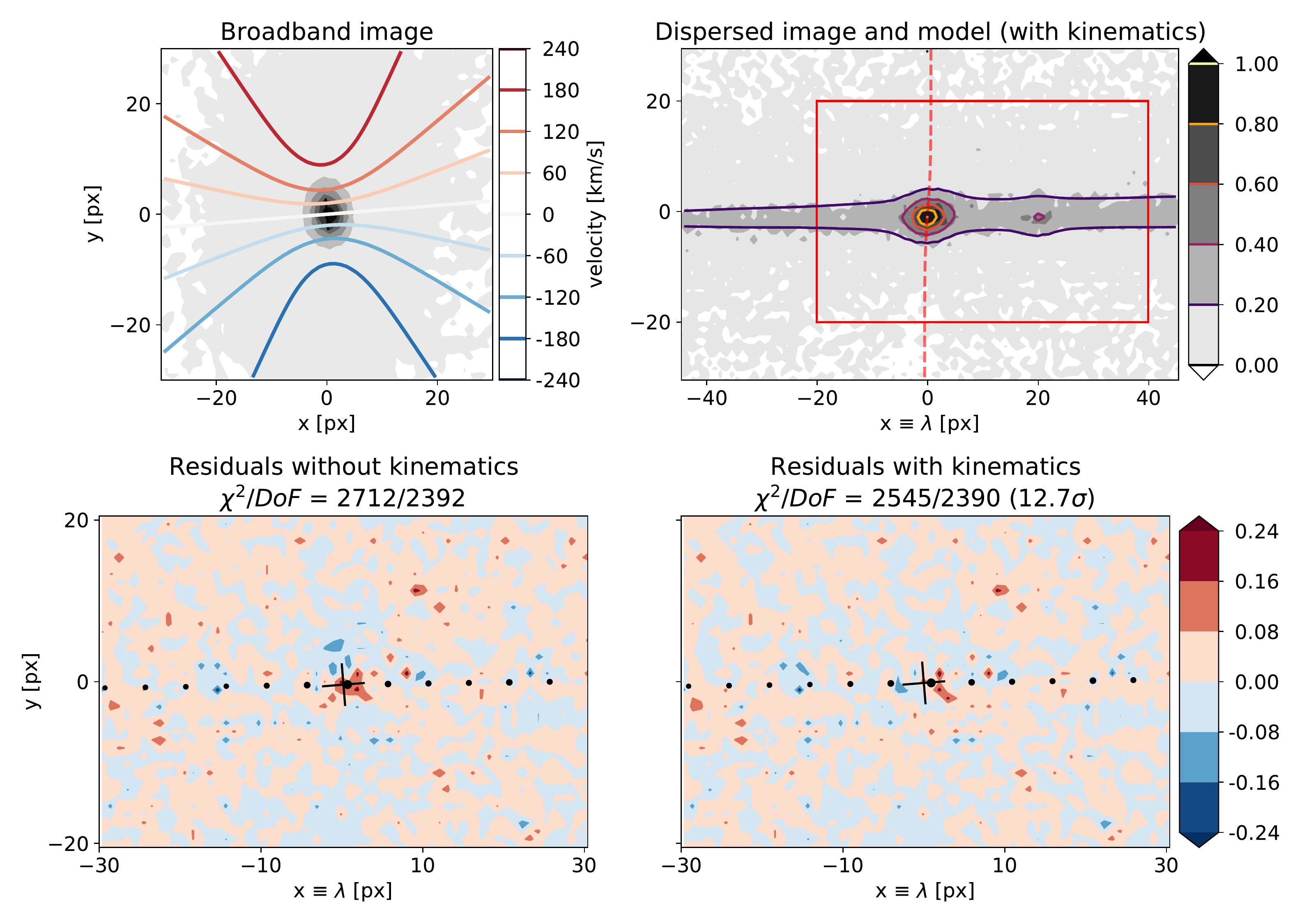}
  \caption{\emph{Top left:} adjusted velocity field $v(\vec{r})$ (contours)
    over-imposed on broadband image $B(\vec{r})$ (gray) of galaxy \#1134 from
    the GLASS survey. \emph{Top right:} input observed (gray) and modeled
    (contours) peak-normalized spectrogram, centered on the
    \Ha{}+[\ion{N}{II}]+[\ion{S}{II}] complex. The \emph{red dashed lines}
    represents the adjusted rotation curve at the \Ha{}
    position. Given the poor kinematic sampling, the offset induced by the
    kinematics is only $\pm \sim0.5$~px. \emph{Bottom left:} residual map for
    the model without kinematics.  \emph{Bottom right:} residual map after
    adding kinematic parameters to the model.  The \emph{black cross} represents
    the position angle of the galaxy at \Ha{} emission line position and the
    \emph{dotted line} the spectral trace. The decrease in the $\chi^{2}$
    corresponds to a $12.7\sigma$ kinematic signature detection.}
  \label{fig:1134_residus}
\end{figure*}

\paragraph{Single line from single roll angle.}
We present the result of the fit performed on galaxy \#1134 from the GLASS
survey, illustrating a very significant kinematic detection. This galaxy has a
strong emission complex \Ha{}+[\ion{N}{II}]+[\ion{S}{II}] and a disc scale
length $r_{d} = \ang{;;0.45}$.  The fit residuals without (resp.\ with)
kinematics is shown in the bottom left (resp.\ right) panel of
Fig.~\ref{fig:1134_residus}.  We note that the residual map without kinematics
displays a coherent structure around \Ha{} line position, which is
significantly reduced in the residual map with kinematics; this indicates that
this structure was a signature of the velocity field.  The fit gives the
following results for the kinematic parameters:
$v_{0}\sin i=\SI{205 \pm 28}{km.s^{-1}}$,
$w_{0}\sin i=\SI{242 \pm 28}{km.s^{-1}.arcsec^{-1}}$ which corresponds to a RC
turnover radius $r_{0} = \ang{;;0.84} \simeq 2 r_{d} \approx \SI{6}{px}$. The
$\chi^{2}$ improvement between the two models (without and with kinematics)
corresponds to a $12.7\,\sigma$~detection ($p \sim \num{e-33}$) of the
kinematic signature.

\begin{figure}
  \includegraphics[width=\columnwidth]{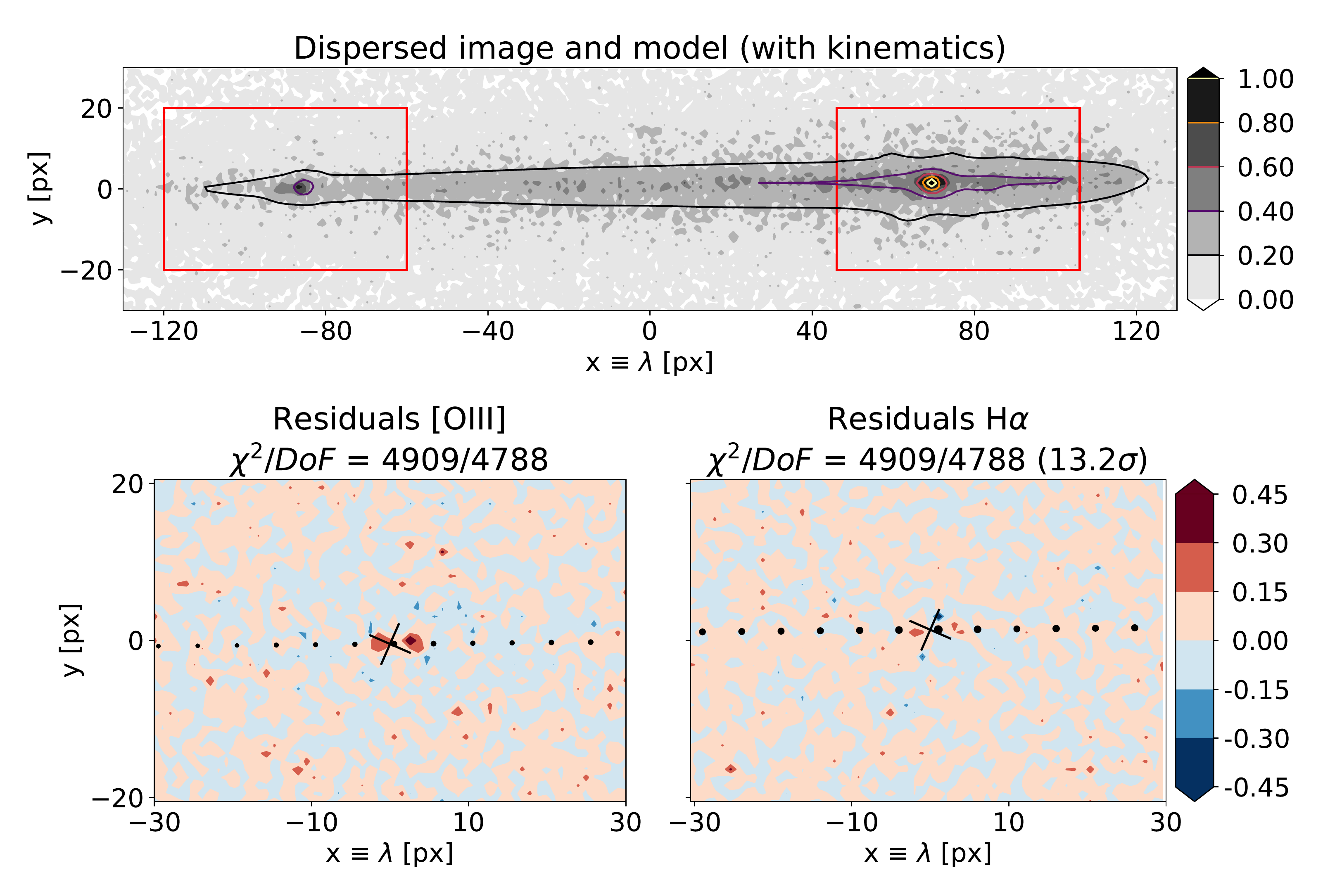}
  \caption{\emph{Top:} input observed (gray) and modeled (contours)
    peak-normalized spectrogram of galaxy \#451 from the GLASS survey for which
    \Ha{} complex (\emph{right frame}) and [\ion{O}{III}] doublet (\emph{left
      frame}) are simultaneously adjusted. \emph{Bottom:} residual map for the
    model with kinematics centered on the [\ion{O}{III}] doublet (\emph{left})
    and on the \Ha{}+[\ion{N}{II}]+[\ion{S}{II}] complex (\emph{right}). The
    \emph{black crosses} represent the PA of the galaxy at [\ion{O}{III}] and
    \Ha{} emission line positions.  The decrease in the total $\chi^{2}$
    corresponds to a joint detection at $13.2\sigma$.}
  \label{fig:451_residus}
\end{figure}

\paragraph{Multiple lines from single roll angle.}
In the case where both \Ha{} and [\ion{O}{III}] lines are detected in
the spectrogram, we can apply a joint fit on both emission line
regions by minimizing the total
$\chi^{2} = \chi^{2}(\Ha{}) + \chi^{2}(\ion{O}{III})$, where each
contribution is computed from its own region.  Both components share
the same intrinsic parameters (redshift, $v_{0}\sin i$, etc.), with
the exception of the continuum level $\mathcal{C}_{0}$ and
cross-dispersion offset $\Delta y$ which can be different.
We show an example of this approach on galaxy \#451 from the GLASS
survey.  The fit residuals with kinematics on each emission line
region is shown in Fig.~\ref{fig:451_residus}.

\paragraph{Multiple roll angles.}
As noticed earlier, each galaxy in the GLASS survey was observed with two
almost orthogonal satellite roll angles, corresponding to different dispersion
directions and therefore relative PAs.  Parameters describing its intrinsic
spectrum, flux distribution and velocity field are common to both spectrograms:
only the transverse offsets $\Delta y$ of the dispersion law may be different.
As a matter of fact, we also observed inconsistencies between roll angles in
the longitudinal component of the dispersion law (i.e.\ wavelength solution),
leading to spuriously different redshifts when estimated from individual
spectrograms.  To account for this effect, we introduced another nuisance
parameter, the wavelength solution offset $\Delta x$ (in \AA) such that the
observed wavelength of an emission line may differ in both roll angles:
\begin{equation}
  \label{eq:z_2PA}
  \lambda = (1 + z) \lambda_{0} \pm \Delta x / 2
\end{equation}
where $z$ is the joint (effective) redshift and $\lambda_{0}$ is the
restframe wavelength.

We are now able to construct a model describing both spectrograms and fit them
simultaneously by minimizing the joint
$\chi^{2} = \chi^{2}(PA_{1}) + \chi^{2}(PA_{2})$, where each $\chi^{2}(PA_{i})$
is the objective function defined in Sec.~\ref{sec:adjust-procedure} for an
individual spectrogram with given relative $PA_{i}$.

\begin{figure*}
  \includegraphics[width=\columnwidth]{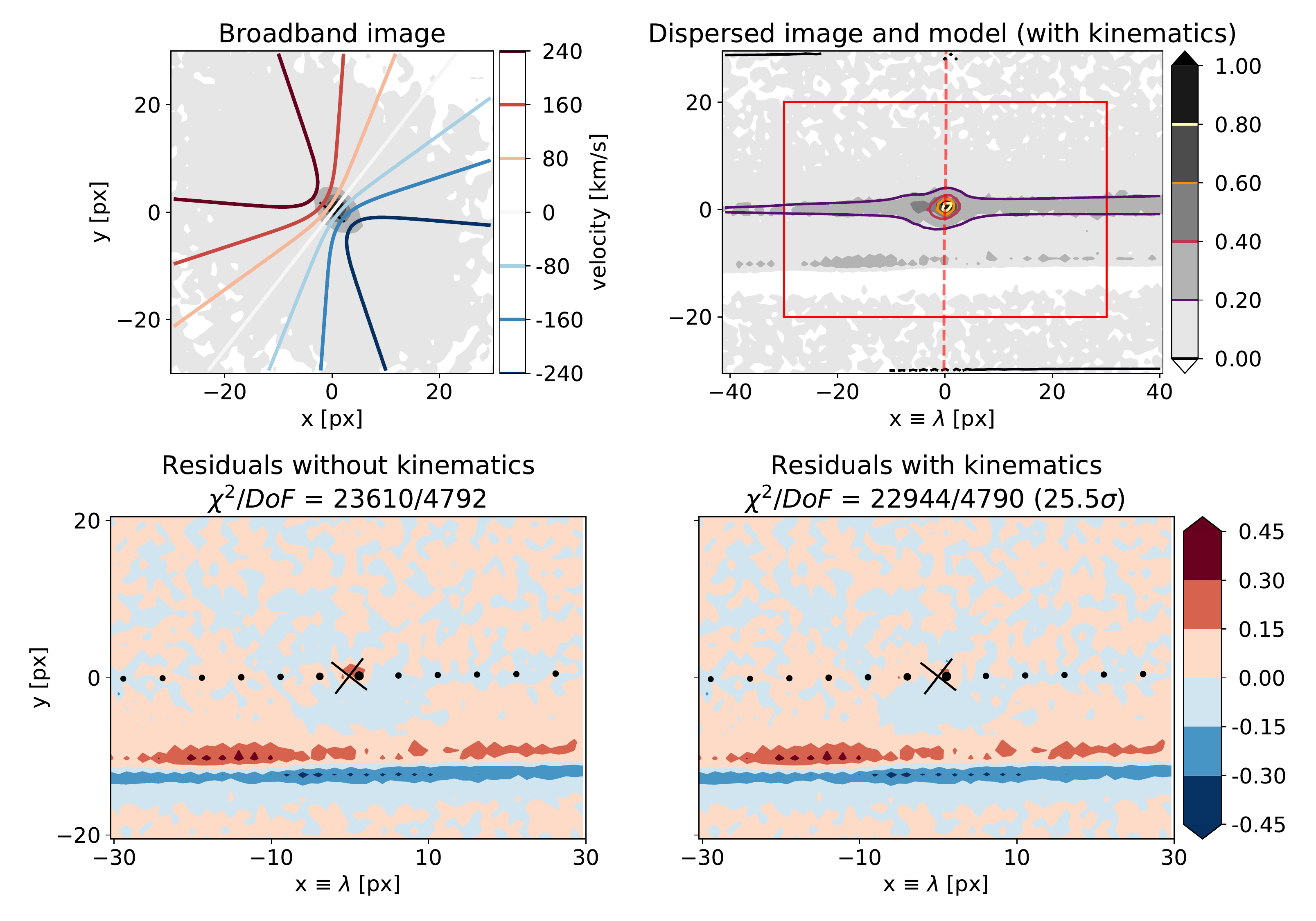}
  \includegraphics[width=\columnwidth]{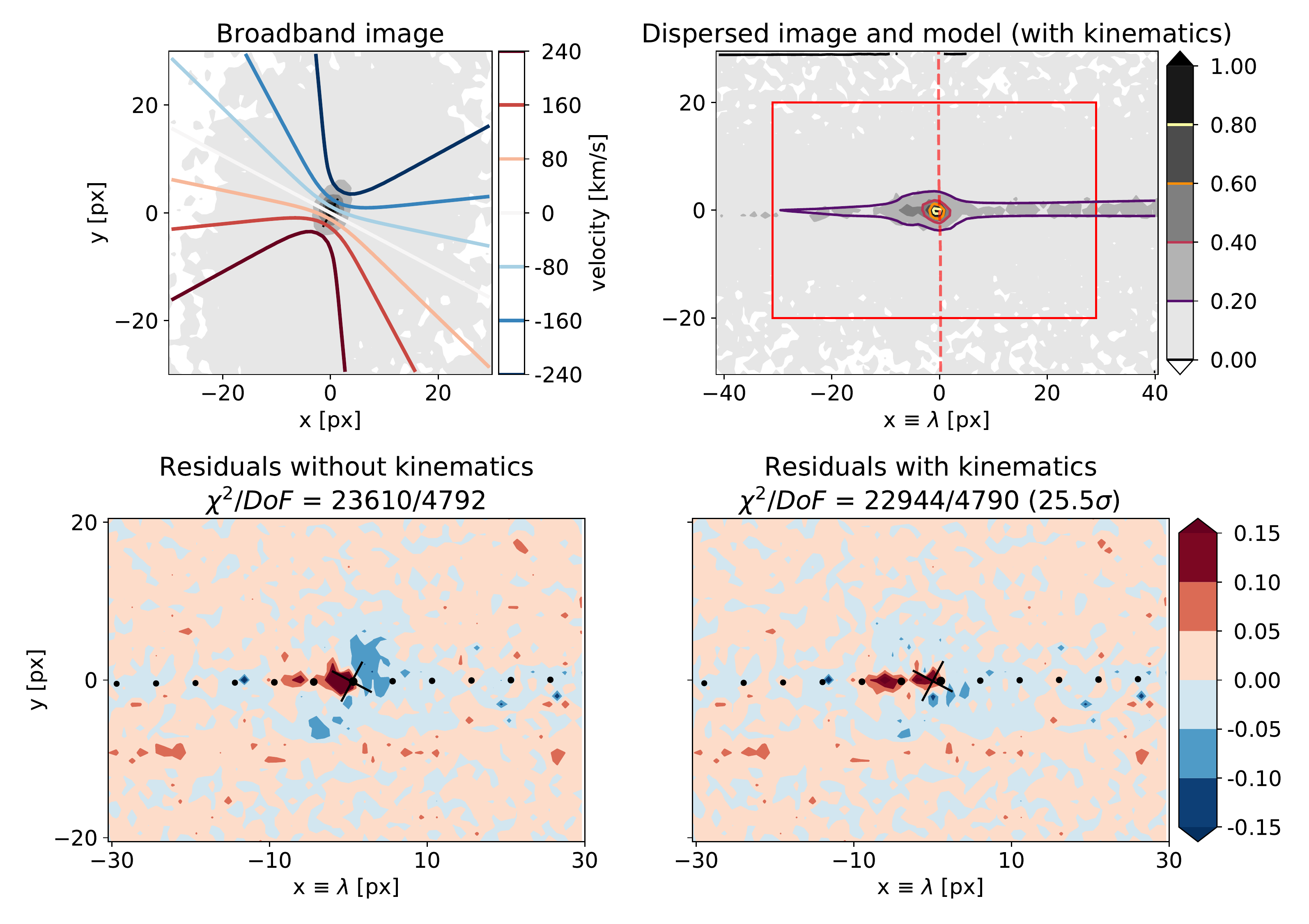}
  \caption{Same as Fig.~\ref{fig:1134_residus} for a joint fit on
    [\ion{O}{III}] emission line of galaxy \#399 from GLASS survey observed
    with a roll angle of $\ang{52}$ (\emph{left panels}) and $\ang{152}$
    (\emph{right panels}).  We note that for the first spectrogram, the large
    residual structure which is visible at the bottom results from an
    incomplete source decontamination: it does not significantly affect the
    model adjustment result, but only increases the final $\chi^{2}$ value. The
    decrease in the joint $\chi^{2}$ corresponds to a $25.5\;\sigma$-level
    detection.}
  \label{fig:399_residus}
\end{figure*}

We apply this simultaneous fit on the [\ion{O}{III}] emission line of
galaxy \#399 from the cluster MACS0717 of the GLASS survey, measured
with a disc scale length $r_{d} = \ang{;;0.44}$.  Each roll-angle
spectrogram is first adjusted on its own, leading to a $14.7\;\sigma$
and $21.6\;\sigma$ internal kinematic detection respectively; the
joint fit on the two spectrograms provides a kinematic $z$-score of
$25.5\;\sigma$ (see Fig.~\ref{fig:399_residus}).
The fit gives the following results for the common adjusted kinematic
parameters: $v_{0}\sin i=\SI{287 \pm 15}{km.s^{-1}}$,
$w_{0}\sin i=\SI{605 \pm 32}{km.s^{-1}.arcsec^{-1}}$ which corresponds
to a RC turnover radius $r_{0} = \ang{;;0.47} \simeq r_{d}$.  We find
a final redshift $z = \num{1.69175 \pm 0.00006}$, accounting for the
wavelength offset $\Delta x = \SI{-12.37 \pm 0.32}{\AA}$; this
corresponds to a shift of about $\pm \sim 0.5$~px between the
wavelength solutions.  It is therefore crucial to include it in the
model to constrain subtle sub-pixel effects such as kinematics from
multi-roll angle spectrograms.

\subsubsection{3D-HST results}
\label{sec:3d-hst-result}

\begin{figure}
  \includegraphics[width=\columnwidth]{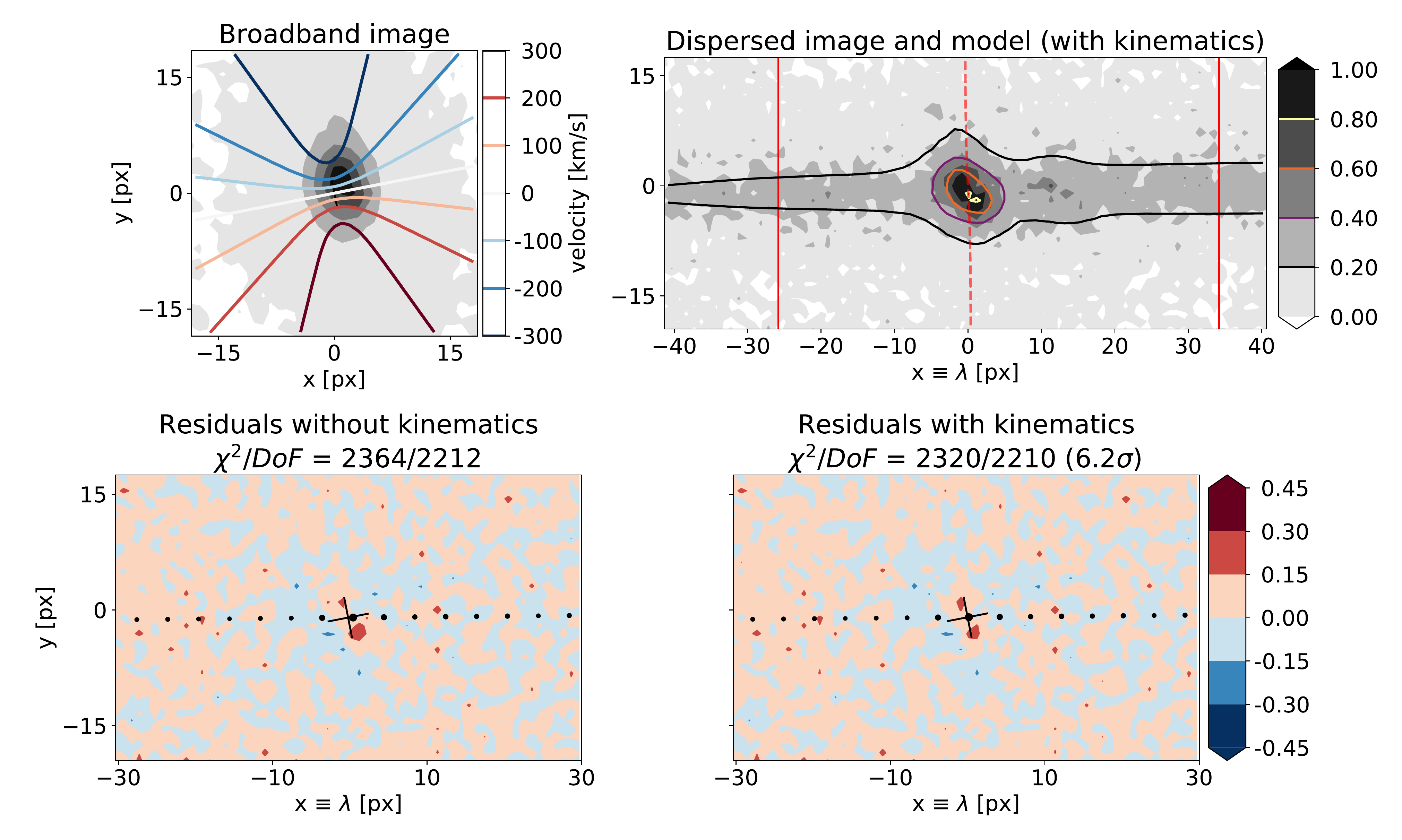}
  \caption{Same as Fig.~\ref{fig:1134_residus} for galaxy \#19843 from the
    3D-HST survey.  The addition of kinematic parameters in the model decreases
    the objective $\chi^{2}$ at a limited $6.2\sigma$~level.}
  \label{fig:19843_residus}
\end{figure}

We present the result for galaxy \#19843 from the 3D-HST survey,
illustrating a weaker kinematic detection than previously shown. This
galaxy has a strong \Ha{}+[\ion{N}{II}]+[\ion{S}{II}] emission complex
and a disc scale length $r_{d} = \ang{;;0.48}$.  As shown in
Fig.~\ref{fig:19843_residus}, the kinematic detection is only at the
$6.2\sigma$~level ($p \sim \num{2e-10}$).

\section{Discussions}
\label{sec:discussions}

\subsection{Core model assumptions}
\label{sec:model-core-assumptions}

\paragraph{Axisymetric thin cold disk.}
Under the cold rotating thin disc hypothesis, morphological and kinematic
position angles are assumed to be the same.  However, the misalignment between
these two position angles can be quite significant ($> \ang{30}$).
\cite{contini_deep_2016} and \cite{harrison_kmos_2017} estimated that such
misalignment can be attributed to either low/high inclinaison, dispersion
dominated systems or complex morphological substructures, such as central bars,
spiral arms or clumps in a low-surface brightness disc.

As the CVG $w_{0}$ is strongly correlated with the assumed kinematic axis (see
Eq.~\eqref{eq:PA_w0}), our method is particularly sensitive to the presence of
central asymmetric structures, such as bars and other kinematically distinct
components (KDC), incompatible with the cold thin disk approximation.  The
method presented here therefore remains best suited to rotation-dominated
isolated late-type spiral galaxies.

\paragraph{Separability assumption.}
The second strong hypothesis of our model is more of a technical one helping to
construct the three-dimensional galaxy datacube; assuming the rest-frame galaxy
spectrum to be uniform over its whole extent. This assumption is acceptable for
homogeneous objects but less valid for KDCs or star-forming galaxies.
\cite{nelson_spatially_2012} studied a sample of 57 galaxies from the 3D-HST
survey and compared \Ha{} and stellar continuum maps: for high-redshift galaxies
($z \sim 1$), \Ha{} emission globally follows the rest-frame $R$-band light but
tends to be more extended and clumpier. Comparing \Ha{} and continuum effective
radii, they found that
$\left\langle r_{e}(\Ha) / r_{e}(R)\right\rangle = 1.3 \pm 0.1$. However, this
difference has not been confirmed with IFU observations for galaxies at
$z \sim 2$ \citep{forster_schreiber_constraints_2011}.

For our study, we have selected only reasonably compact objects, and
systematically found $\eta \sim 1 \pm 0.1$, indicating that \Ha{} and
continuum flux distribution are reasonably similar.

Overall, the two core assumptions are well justified for large and bright
isolated late-type galaxies. Furthermore, given the available instrumental
setup, with limited spatial and kinematic samplings, slitless observations can
only constrain simplified models (disk rotation curves, age or metallicity
gradients, etc.).

\subsection{Systematics from the position angle}
\label{sec:PA-systematic}

As mentioned in Sec.~\ref{sec:resolv-kinem-slitl}, galaxy relative PA and CVG
$w_{0}$ induce similar line shape distortions in the spectrogram, and are
therefore strongly correlated.  By differentiating Eq.~\eqref{eq:PA_w0} at
constant effective PA, one can estimate the systematic error $\Delta w_{0}$
associated to an error on the relative PA of the galaxy:
\begin{equation}
  \label{eq:PA_w0_diff}
  \frac{\Delta w_{0}}{\Delta PA} = \frac{R_{kin}/s + w_{0} \sin PA}{\cos PA}.
\end{equation}
For the galaxy \#1134 from the GLASS survey, we find a systematic
error on the CVG $\Delta w_{0} \approx \SI{38}{km.s^{-1}.arcsec^{-1}}$.
For both the galaxies \#451 and \#399, $\Delta w_{0} \approx
\SI{123}{km.s^{-1}.arcsec^{-1}}$. Finally, for the galaxy \#19843 of the
3D-HST survey, we compute $\Delta w_{0} \approx
\SI{108}{km.s^{-1}.arcsec^{-1}}$.

Overall, the systematic uncertainties on the CVG $w_{0}$ are
larger than the statistical ones (except for galaxy \#19843) but remains
in a similar range since the $PA$ is very well constrained from the
broadband images (see Table~\ref{tab:redshift-fit-precision}) and  the
kinematic sampling is the governing term in Eq.~\eqref{eq:PA_w0_diff}
where $s=\SI{0.13}{arcsec.px^{-1}}$.

\subsection{Implications on redshift precision}
\label{sec:implication-redshift-precision}

Self-confusion effect in slitless spectroscopy is generally assumed to induce
redshift measurements less accurate than in more traditional spectroscopic
observations.  However, the forward model presented here significantly
attenuate this effect, as the effective spectral resolution does not depend on
the object extent anymore, \emph{even in the absence of kinematic parameters}.
In Table~\ref{tab:redshift-fit-precision}, we compare the redshift
uncertainties from the forward model described in Sec.~\ref{sec:results} to the
total uncertainties (including systematics) estimated by 3D-HST and GLASS
analysis.  We emphasize that only the \emph{statistical} error on the redshift
measurement is quoted in our case, as we do not have access to the data
reduction details for HST-based surveys.  Nonetheless, we point out that the
forward modeling delivers a redshift accuracy down to $\sim \num{e-4}$
or less. Using a similar forward approach for slitless spectroscopic
reduction would probably provide an equivalent precision gain on
instrumental calibrations (notably dispersion solution), which would in
turn decrease the final systematic errors to a comparable level.
Ultimately, our study suggests that a consistently forward analysis of
slitless observations -- including for data reduction procedures --
could allow a significant gain in redshift precision.

\subsection{Kinematic measurements}
\label{sec:kinematic-measurement}

Notwithstanding its direct degeneracy with the assumed kinematic angle, the CVG
$w_{0}$ is reasonably constrained by the core 2D shape of the emission lines in
the slitless spectrogram.  On the other hand, the plateau velocity $v_{0}$
remains difficult to measure, since spectroscopic observations are rarely deep
enough in available redshift surveys to probe regions well beyond effective
radius $r_{d} \approx r_{0}$ where the velocity flattening would be manifest.
Furthermore, current instrumental setups, with a limited kinematic sampling
($R_{kin} > \SI{650}{km.s^{-1}.px^{-1}}$), do not favor precise measurements of
internal kinematic signatures, inducing sub-px spectral distortions.

We note that in this proof-of-concept analysis, we did not take into account the
``beam smearing'', i.e.\ the degradation of kinematic resolution due to limited
spatial resolution.  Formally, one would need to construct an ``infinitely''
spatially resolved model cube before applying the spatial PSF, while, in our
case, we built the model cube from an already PSF-convolved flux model
(Eq.~\eqref{eq:2}).  This would mostly change the estimated value of $w_{0}$
for very steep unresolved CVGs.

Last, the natural drawback of the forward approach is that the model is
assumed to be a fair representation of the observed data. If this is not the
case (specifically, non cold thin disc-like objects and/or non-radial
structures), errors are dominated by systematics and the resulting kinematic
parameters have no adequate physical interpretations.


\section{Conclusions and perspectives}
\label{sec:conclusion-perspective}

In this article, we explored the possibility to probe single object with
slitless spectroscopy by measuring not only integrated spectral features but
also spatially resolved quantities such as internal kinematics. To achieve
that, we presented a forward-model of slitless spectrograms from a galaxy model
-- including flux distribution, intrinsic spectrum and kinematic parameters --
and an instrumental signature. This method relies on two major assumptions: the
axi-symmetric thin cold disk approximation and the separability hypothesis.  We
applied this method on HST-surveys galaxies to measure internal kinematics
parameters: the plateau velocity $v_{0}$ and the central velocity
gradient~$w_{0}$.

The kinematic signature on slitless spectrograms is very specific, as a
quadrupole structure in the kinematic-less residuals around emission lines.  It
is therefore difficult to confuse with other effects such as line flux
distribution or radial structures.  Even if the kinematic signature is subtle,
it extends on the full galaxy scale, which makes the statistical detection
significant even with limited kinematic sampling.

The mains results of our method can be summarized as follows:
\begin{itemize}
\item We present the first detection of resolved internal kinematics in
  galaxies from slitless spectroscopic observations.  Using a simple asymptotic
  velocity curve, we note that the CVG $w_{0}$ is more precisely determined
  than the plateau velocity $v_{0}$, but is directly sensitive to systematic
  errors on the assumed relative position angle of the galaxy, which must be
  estimated from external photometry.

\item The forward model naturally corrects for the self-confusion effect: the
  spectral resolution does not explicitly depend on the object shape/extent
  anymore.  It implies a more precise redshift measurement (down to \num{e-4})
  with a simple model without kinematics and even better with a model including
  kinematics.  Improved redshifts from slitless surveys will have a direct
  impact on cosmological probes as Baryon Acoustic Oscillations and Redshift
  Space Distortions measurements, where the expected precision is
  \numrange{e-3}{e-4}.

\item We observe sub-px inconsistencies in the dispersion law of the
  spectrograms from 3D-HST and GLASS surveys in the cross-dispersion direction,
  and use a nuisance parameter to account for it.  However, calibration errors
  along the dispersion axis induce a systematic error in the effective redshift
  determination, but do not affect the kinematic parameters. It should
  be kept in mind that this analysis is a proof of concept applied at the
  data analysis level only: optimal results would be obtained with a full
  forward model of slitless spectrograms, including in data calibration
  stages.

\item We demonstrate the great flexibility of our forward model, where one can
  simultaneously fit different emission lines, spectrograms observed with
  different dispersion direction, and even spectrograms from different
  instruments, allowing improved constraints on redshift and kinematic
  parameters.
\end{itemize}

The spectroscopic resolving power $\mathcal{R}$ is a key parameter determining
the amplitude of the Doppler distortion in the spectrogram and therefore is a
deciding quantity to constrains precisely the CVG and the plateau velocity.
Furthermore, the spectroscopic observations must be deep enough to probe the
velocity curve well beyond the photometric disk radius and reach the asymptotic
regime.

We stress out that, building on a finer understanding of the
spectrogram formation, this forward model of slitless spectra has
the potential to obtain more precise redshift measurements than standard
approaches.  This will be massively tested on futures slitless
surveys such as Euclid which will acquire spectrograms for 30M
galaxies in the \emph{Wide Field} and will also produces deeper spectroscopic
observations in the \emph{Deep Fields} \citep{laureijs_euclid_2011}. Moreover,
these surveys will be characterized by a finer kinematic sampling
$R_{kin} \sim \SI{200}{km.s^{-1}.px^{-1}}$, more suited for kinematic
measurements and galaxy scaling relation studies such as the Tully-Fisher
relation \cite[e.g.][]{aquino-ortiz_kinematic_2018, lelli_baryonic_2019} or
morpho-kinematic classification \cite[e.g.][]{cortese_sami_2014,
  kalinova_towards_2017, schulze_kinematics_2018}.

\begin{acknowledgements}
  We thank the referee for the careful reading of the paper and useful
  suggestions.  We also thank Orane Monteil for an early contribution to this
  work. This research made use of the MINUIT algorithm
  \citep{1975CoPhC..10..343J} via the
  \texttt{iminuit}\footnote{\url{https://github.com/iminuit/iminuit}}
  Python interface. This work is based on observations taken by the 3D-HST
  Treasury Program (GO 12177 and 12328) with the NASA/ESA HST, which is
  operated by the Association of Universities for Research in Astronomy, Inc.,
  under NASA contract NAS5-26555.
\end{acknowledgements}

%
%

\bibliographystyle{aa} 
\bibliography{aa} 

\end{document}